\documentclass[sigconf]{acmart}

\AtBeginDocument{%
  }

\setcopyright{acmlicensed}
\copyrightyear{2025}
\acmYear{2025}
\acmDOI{XXXXXXX.XXXXXXX}

\acmConference[Arvix]{preprint arXiv}{Oct}{2025}

\usepackage{algorithm}
\usepackage{algpseudocode}
\usepackage{subcaption}
\usepackage{pifont}

\newtheorem{definition}{Definition}

\newcommand{\algo}{CSV}
\begin{document}

\title{Learned Indexes with Distribution Smoothing via Virtual Points}

\author{Kasun Amarasinghe}
\affiliation{%
  \institution{The University of Melbourne}
  \city{Melbourne}
  \country{Australia}
  }
\email{kasun.amarasinghe@student.unimelb.edu.au}

\author{Farhana Choudhury}
\affiliation{%
  \institution{The University of Melbourne}
  \city{Melbourne}
  \country{Australia}
  }
\email{farhana.choudhury@unimelb.edu.au}

\author{Jianzhong Qi}
\affiliation{%
  \institution{The University of Melbourne}
  \city{Melbourne}
  \country{Australia}
  }
\email{jianzhong.qi@unimelb.edu.au}

\author{James Bailey}
\affiliation{%
  \institution{The University of Melbourne}
  \city{Melbourne}
  \country{Australia}
  }
\email{baileyj@unimelb.edu.au}

\begin{abstract}
  Recent research on learned indexes has created a new perspective for indexes as models that map keys to their respective storage locations. These learned indexes are created to approximate the cumulative distribution function of the key set, where using only a single model may have limited accuracy. To overcome this limitation, a typical method is to use multiple models, arranged in a hierarchical manner, where the query performance depends on two aspects: (i) traversal time to find the correct model and (ii) search time to find the key in the selected model. Such a method may cause some key space regions that are difficult to model to be placed at deeper levels in the hierarchy. To address this issue, we propose an alternative method that modifies the key space as opposed to any structural or model modifications. This is achieved through making the key set more learnable (i.e., smoothing the distribution) by inserting virtual points. Furthermore, we develop an algorithm named \algo\ to integrate our virtual point insertion method into existing learned indexes, reducing both their traversal and search time. We implement \algo\ on state-of-the-art learned indexes and evaluate them on real-world datasets. Extensive experimental results show significant query performance improvement for the keys in deeper levels of the index structures at a low storage cost.
\end{abstract}



\maketitle

\section{Introduction}
Learned indexes~\cite{kraska2018case} have reported strong query performance and are attracting much attention from both the academia and industry in recent years. 
The core idea of learned indexes is that an index structure can be seen as a mapping function $f(\cdot)$ from a search key $k_i$ to the storage location (i.e., the rank $rank(k_i)$) of the corresponding data record: $rank(k_i)\approx f(k_i)$.  The mapping function (a.k.a. \emph{indexing function}) is learned and approximated by machine learning algorithms (models). 
To enable the learning, a storage ordering needs to be established. Typically, an ascending order based on the search keys is used, such that the mapping function is effectively the \emph{cumulative distribution function} (CDF) of the search keys. 

\begin{figure}[t]
  \centering
  \begin{subfigure}[b]{0.4\textwidth}
    \includegraphics[width=\textwidth]{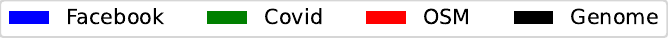}
  \end{subfigure}

  \begin{subfigure}[b]{0.27\textwidth}
    \includegraphics[width=\textwidth]{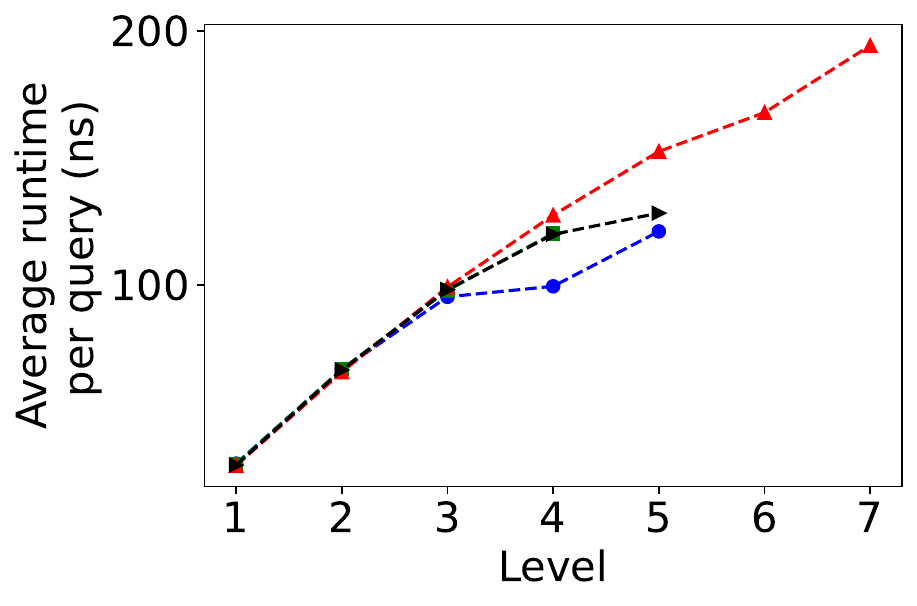}
    \end{subfigure}
    \caption{Query time at each level of the LIPP index for four real datasets, each with 200 million keys.}
 \label{fig:query_levels_lipp}
\end{figure}

Different learned indexes have been proposed~\cite{ding2020alex, ferragina2020pgm,wu2021lipp, zhang2021colin,nathan2020flood, li2020lisa, ding2020tsunami, ding2022hybrid_error, pai2023workloadz, sheng2023wisk, li2021finedex, tang2020xindex, ge2023sali, lu2021apex, lan2023aulid, wang2024wipe}, with a common theme to design indexing functions and structures that enable better approximation of the CDF, since approximation errors translate to mapping errors (i.e., difference between $rank(k_i)$ and $f(k_i)$) and hence extra search costs to examine data records at around $f(k_i)$ and recover from the errors.
However, such approaches mean the use of either complex indexing functions (e.g., splines~\cite{kipf2020radixspline, mishra2021rusli}) or piece-wise functions with many segments~\cite{ferragina2020pgm, li2021finedex, galakatos2019fiting}, both of which could lead to sub-optimal query efficiency. This is illustrated by Fig.~\ref{fig:query_levels_lipp} with LIPP~\cite{wu2021lipp} -- one of the latest learned indexes. The index has a hierarchical structure built in a top-down manner. When an index model (i.e., a node in the index) cannot achieve an overall low mapping error for all keys assigned to it for indexing, sub-index models are created recursively as a hierarchy, to accommodate for the ``more difficult to learn'' keys. 
As Fig.~\ref{fig:query_levels_lipp} shows, keys indexed in deeper levels (higher levels in the figure) reported higher query times on average, on all four datasets. 

In this paper, we approach the problem from an alternate perspective -- we adjust the CDF such that it becomes easier to be approximated by the indexing functions, to achieve lower approximation errors and higher query efficiency. 

\begin{figure}[t]
  \centering
  \begin{subfigure}[b]{0.23\textwidth}
    \includegraphics[width=\textwidth]{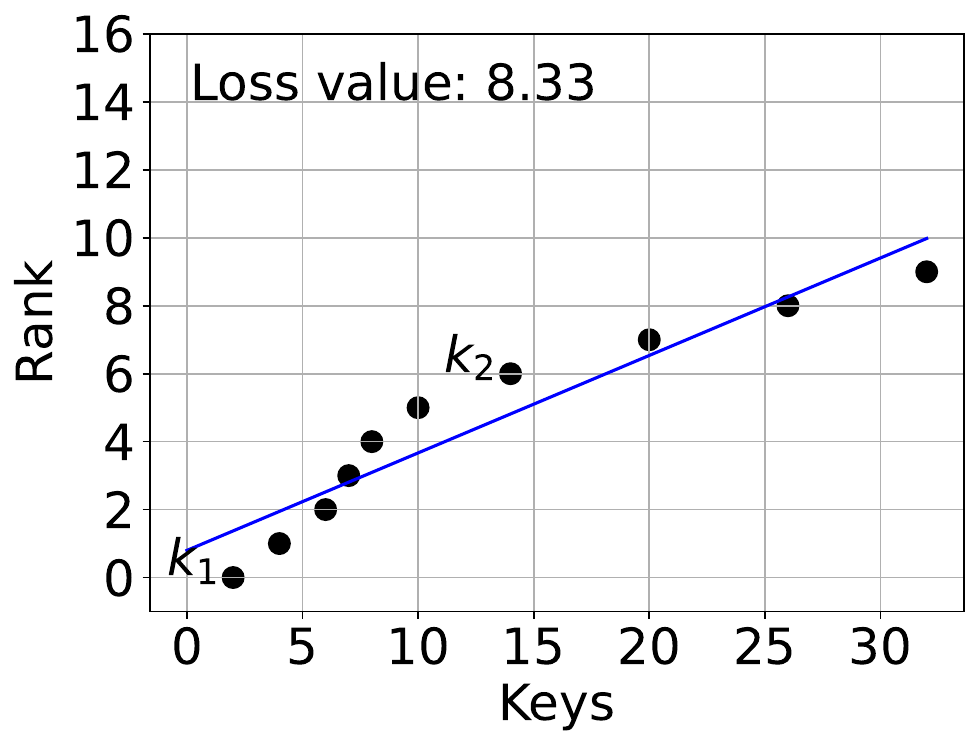}
    \caption{Before smoothing}
    \label{fig:before_smooth}
  \end{subfigure}
  \hfill
  \begin{subfigure}[b]{0.23\textwidth}
    \includegraphics[width=\textwidth]{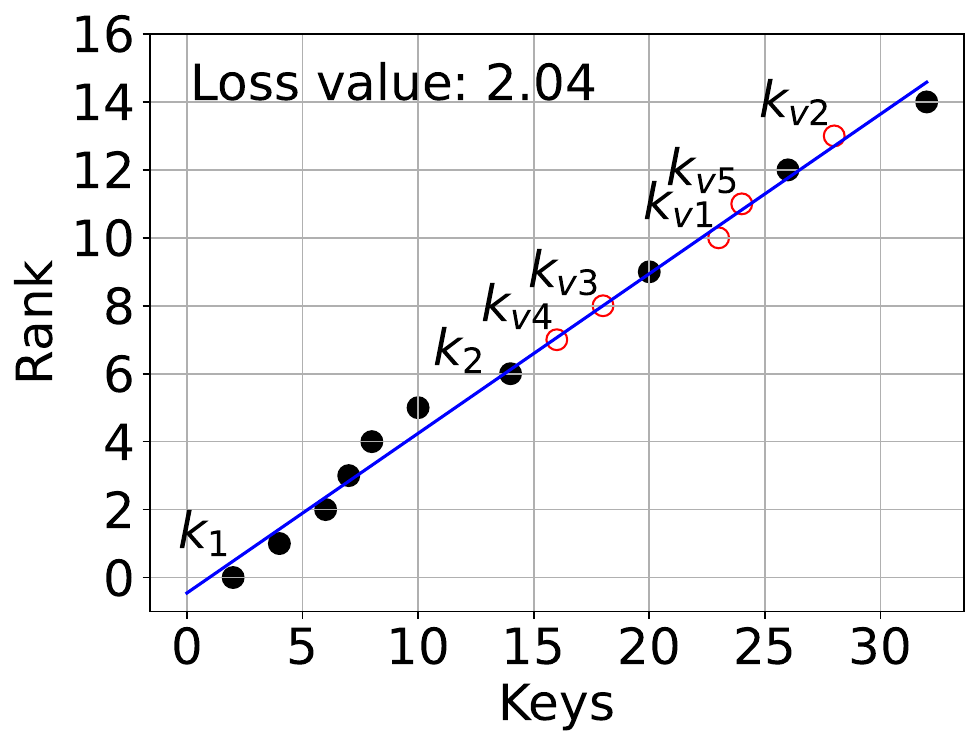}
    \caption{After smoothing}
    \label{fig:after_smooth}
  \end{subfigure}
  \caption{Indexing data points (keys) with CDF smoothing.}
    \label{fig:scatter_keys}
\end{figure}

Our core idea is to add \emph{virtual points} to ``smooth'' the CDF of a dataset. Take Fig.~\ref{fig:before_smooth} as an example, where each black dot represents a data point (i.e., its search key). Approximating the CDF of the dataset with a linear function can result in a large approximation error (and hence high search costs at query time) for keys $k_1$ and $k_2$. We sum up the squared prediction error of every point:  
\begin{equation}
\mathcal{L}_f(K) = \sum_{i = 1}^{n} \big(f(k_i) - rank(k_i)\big)^2, 
\label{eq:sse_ori}
\end{equation}
where $K$ denotes the set of keys and $n$ is its size, $k_i \in K$ is a key, $rank(k_i)$ is its rank, and $f(\cdot)$ is the indexing function. 
We refer to $\mathcal{L}_f(K)$ as the \emph{loss function}, for which we use the \emph{sum of squared errors} (SSE). In this case, 
$\mathcal{L}_f(K) = 8.33$ -- a large value of $\mathcal{L}_f(K)$ suggests worse prediction accuracy using $f$ for search key mapping and hence higher query times. 

As Fig.~\ref{fig:after_smooth} shows, we can add  virtual points $V = \{k_{v1}, k_{v2}, \ldots, k_{v5}\}$ represented by the red hollow dots. Here, we assume a \emph{smoothing budget} of $0.5n = 5$, i.e., 5 virtual points are allowed. 
Now the original data points are spread out, and the CDF of the (original and virtual) points is closer to a straight line. We refit the points with a new indexing function $f'$, with the loss function value $\mathcal{L}_{f'}(K)$ being reduced to 2.04 (and $\mathcal{L}_{f'}(K\cup V)$ = 2.29). 

We show that, given a smoothing budget that constrains the additional space costs, finding the optimal placement of the virtual points to minimise the loss function  $\mathcal{L}_{f'}(K)$ 
is NP-hard. We then propose approximation solutions for two generic scenarios: (1) smoothing the CDF for index learning with a single indexing function and (2)~smoothing the CDF for a hierarchy of indexing functions, which is a common structure for existing learned indexes.  
We focus on linear indexing functions for their  efficiency, although CDF smoothing can naturally extend to more complex (e.g., quadratic) functions.

We propose an algorithm named \emph{\underline{C}DF \underline{s}moothing via \underline{v}irtual points} (\emph{\algo}) to smooth the CDF for optimising hierarchical learned indexes, with the aim to reduce the overall height of the structures as well as the prediction errors of each indexing function, and hence the query costs. This is performed by collecting sub-trees of the hierarchical structure and smoothing the CDF of the keys in them. As a result, the keys could now be placed into a single node due to the higher learnability, provided it surpasses a cost model threshold value. Here, the cost model is used to balance the reduction in index traversal time and the potential increase in the leaf-node search time due to the increase of keys in a node.

It is important to note that our aim is \emph{not} to propose yet another learned index but rather a technique that can be integrated with existing or emerging hierarchical learned indexes to optimise their query efficiency with controllable extra space. To show the applicability of our \algo\ algorithm, we integrate it with three recent learned indexes ALEX~\cite{ding2020alex}, LIPP~\cite{wu2021lipp}, and SALI~\cite{ge2023sali}, which are the state-of-the-art (SOTA).  

To summarize, this paper makes the following contributions:

(1)~We propose a key space transformation technique using CDF smoothing via inserting virtual points to enhance index learnability.

(2)~We propose an efficient algorithm named \algo\ to integrate the CDF smoothing technique with hierarchical learned indexes to improve the query performance, with a controllable space overhead.

(3)~We integrate \algo\ with three learned indexes ALEX, LIPP, and SALI, and we conduct experiments with four real datasets. The experimental results show that the learned indexes powered by \algo\ manage to promote up to 60\% of the keys in lower levels to upper levels, resulting in up to 34\% improvement of their query time, with less than 15\% increase to the storage space overhead. 

\section{Related Work}
We first review learned indexes in general. Then, we focus on studies addressing complex distributions, which share a similar goal with us. We also cover a technique called the poisoning attacks, which motivates our CDF smoothing technique.

\subsection{Learned Indexes}\label{sec:learned_index} 
Learned indexes are a trending topic in the database community~\cite{ding2020alex, ferragina2020pgm, wu2021lipp, zhang2021colin,ding2022hybrid_error,kraska2018case,kipf2020radixspline,lu2021apex,yang2023flirt, lan2023aulid}. 
Their key idea is to treat indexes as functions that map a search key to the storage position of the corresponding data object, which can be learned with machine learning models. A common approach is to lay out the data objects by ascending order of their search keys, such that the indexing functions are effectively (approximations of) CDFs of the keys. 

To index large datasets, multiple indexing functions are used, typically organized in a hierarchical structure like a B-tree. 
The lookup performance of such a structure is then dominated by two steps: 
(1)~the \emph{traversal time} to find the leaf-node (every node corresponds to an indexing function) indexing the search key, and (2)~the search time within the selected leaf-node (\emph{leaf-node search time} hereafter to distinguish from the traversal time) to locate the target data object, as the indexing functions have errors and may not produce the exact storage position of the search target~\cite{sun2023learned, wongkham2022updatable}.

It is a challenge to balance the query costs from the two steps above. While a deeper structure with more indexing functions may fit the data distribution better and have lower leaf-node search times, it may also have higher traversal times and larger index sizes~\cite{wu2022nfl,ge2023cutting}. 
Some studies impose a maximum error bound on the indexing functions to reduce the leaf-node search times~\cite{galakatos2019fiting,ferragina2020pgm}, also at the cost of more indexing functions. 
Another approach is to use more complex indexing functions (as opposed to linear ones)~\cite{kraska2018case,tang2020xindex,kipf2020radixspline}, e.g., splines, which could better fit the CDFs. The issue with this approach is the higher inference time for the function, and hence higher query and insertion times~\cite{sun2023learned}. 

These studies design structures and indexing functions to better fit the data distribution. We address the challenge from an alternate perspective, i.e., we adjust the data distribution such that it is easier to be fitted by the indexing functions.  

\subsection{Addressing Complex Data Distributions}\label{sec:learned_index_distribution}
To better index CDFs of complex data distributions, there are two common approaches. One is to use more complex functions such as splines and piece-wise linear regression models \cite{ferragina2020pgm, kipf2020radixspline}. The other is to use better data partitioning strategies for easier CDF learning over each partition, such as by CARMI~\cite{zhang2021carmi} and EWALI~\cite{liu2022ewali}. Another study, LER~\cite{eppert2021tailored}, uses logarithmic error-based loss functions (instead of the more commonly used least squared error-based) to improve the learning of index models that better fit the CDF. 

A latest development, SALI~\cite{ge2023sali}, identifies the most frequently accessed nodes via probability models given a query workload. The corresponding sub-trees are flattened using a segmentation approach, similar to the PGM index~\cite{ferragina2020pgm}, to reduce their traversal time. However, this leads to an additional search step for queries, as we need to find the correct node from the flattened structure. 

A couple of studies~\cite{wu2022nfl,li2021gappluggable} transform the input key set into a more uniform distribution to improve the CDF learnability. The NFL index~\cite{wu2022nfl} transforms the key distribution using a numerical normalizing flow that transforms a latent distribution to a new distribution via generative models. The distribution transformation introduces overheads, while queries also need to be transformed to use the index. Further, the transformation may increase the tail conflict degree for certain distributions, making it unsuitable in those instances. 
The \emph{gap insertion}~\cite{li2021gappluggable} technique inserts gaps between the keys (i.e., storage positions of the corresponding data objects) to straighten the CDF of the keys, thereby improving its learnability. 
However, this is performed by manipulating the rank of each key, and as a result, multiple keys could be given the same position. An extra array is used to house such conflicting keys, which in turn introduces search overheads to locate the correct key.
This method leads to a heavy storage space increase of up to 87\%.

Several learned indexes~\cite{ding2020alex,wu2021lipp,lu2021apex} leave gaps in their storage structure (i.e., gapped arrays). While their purpose is to accommodate data insertions, a side effect is changing the data distribution, which is what we do. A core difference to note is that, they do not consider minimizing the indexes' model prediction errors when adding gaps, in contrast to our approach which does. 

\subsection{Poisoning CDFs} \label{sec:poisoning}
Our idea of adjusting data distribution to fit the indexing functions is rooted from \emph{data poisoning} -- a process of manipulating the training data to change the results from a predictive model~\cite{kornaropoulos2022price}. Data poisoning has been introduced into learned indexes to poison the indexing functions and negatively impact their capability to approximate the CDFs~\cite{kornaropoulos2022price}. The main goal of this process is to identify new points to include into the original key set that would cause the maximum increase to the loss function value (i.e., the SSE). 

Motivated by the poisoning technique, we propose a technique that smooths the data distribution by adding virtual points, to obtain CDFs that are easier to be approximated by indexing functions (models), hence leading to a structure with higher query efficiency. Since the models are built with virtual points that can be used to host data insertions, a side benefit of our structure is that it is more resilient against data insertions. Table~\ref{tab:comp_methods} highlights the key difference between our technique \algo, NFL, and gap insertion (GI). 

\begin{table}[t]
\caption{Comparison with Existing Works}
    \label{tab:comp_methods}
    \centering
\setlength{\tabcolsep}{2pt}
\begin{tabular}{lrrr}
\toprule
           & \textbf{\algo}  & NFL~\cite{wu2022nfl}  & GI~\cite{li2021gappluggable} \\ \midrule
No extra transformation at query time & \ding{51}       & \ding{55}
   & \ding{51}     \\ \hline
Low storage overhead  & \ding{51}  & \ding{51}  & \ding{55} \\  \hline
Integrable into other learned indexes & \ding{51}        & \ding{51}    &  \ding{55} \\  \hline
Robust across different distributions & \ding{51}        & \ding{55}  & \ding{51} \\
\bottomrule
\end{tabular}
\end{table}

\section{Preliminaries}
\textbf{Problem statement.} 
Consider a dataset $D$ of $n$ data records, where each record is associated with a one-dimensional value as its index key. Let $K$ be the list of all index keys associated with $D$, sorted in ascending order. 

Suppose that the index keys have been partitioned and indexed by a set $\mathcal{F}$ of $m$ indexing functions. Each indexing function $f_i \in \mathcal{F}$ has some prediction error (a.k.a ``loss'') for a key $k \in K_i$ indexed by it. Here, $K_i \subset K$ refers to the subset of keys indexed by $f_i$. The loss refers to the squared difference between the predicted index position $f_i(k)$ and the rank of $k$ in $K$, i.e., $rank(k)$. The sum of squared errors (SSE) is one of the most commonly used metric to represent loss in the existing studies of learned indexes. Let $\mathcal{L}_{\mathcal{F}}$ be the \emph{total sum of squared errors} of all indexing functions in $\mathcal{F}$:
\begin{equation}\label{eq:total_error}
\mathcal{L}_\mathcal{F}(K) = \sum_{i = 1}^{m} \sum_{k \in K_i} \big( f_i(k) - rank(k)\big)^{2}.
\end{equation}

Equation~\ref{eq:total_error} is the loss function of our optimisation problem.  
We aim to insert values (\emph{virtual points}) into $K$ while keeping it sorted, such that $\mathcal{L}_\mathcal{F}(K)$ is minimised, i.e., to \emph{smooth the CDF} of $K$. 

A naive optimal smoothing scheme is to insert as many virtual points as needed such that every point $k \in K_i$ lies at the $f_i(k)$-th position (i.e., $rank(k) = f_i(k)$) in the list (assuming unique integer keys). This way, the loss becomes zero after smoothing. 
In reality, this smoothing scheme is infeasible, due to the non-uniqueness of the keys in $K$ and the potentially high space cost. For example, if the keys are 64-bit integers, it will take $2^{64} \times 8$ bytes $\approx 128$ exabyte (i.e., $128 \times 10^6$ TB) to achieve such an index key layout.

To balance between the space overhead and the smoothness of the CDF with inserted points, we consider a ``smoothing budget'' $\lambda$, i.e., the number of virtual points allowed to be inserted, such that the loss is minimised given the constraint of $\lambda$. We assume $\lambda = \alpha \cdot n$ where \emph{the smoothing threshold} $\alpha$ is in $(0, 1)$, to retain a linear space overhead. Formally, we aim to solve the following problem:

\begin{definition}\label{def:problem}
[\textbf{Learned index smoothing}]
Given a list of index keys $K$ sorted in ascending order and partitioned into $m$ segments, each of which is indexed by an indexing function $f_i \in \mathcal{F}$, the learned index smoothing problem aims to insert a set $V$ ($|V| \le \lambda$) of virtual points into $K$ while keeping $K$ in order, such that the loss as defined by Equation~\ref{eq:total_error} is minimised.
\end{definition}

We consider linear indexing functions as they are used in most existing learned indexes. To simplify the discussion, we use integer index keys, while our techniques also apply to real number index keys when they can be scaled up to become integers. 

\textbf{NP-hardness analysis.}
Solving the exact CDF smoothing problem is NP-hard as it can be reduced from the Knapsack problem, which is a known NP-hard problem. 
\begin{lemma}
    Learned index CDF smoothing is NP-hard.
\end{lemma}
\begin{proof}

We reduce from the Knapsack problem which is  NP-hard. 
The Knapsack problem considers a set of items $S$. Each item $s \in S$ has a value $c_s$ and a weight $w_s$. The objective is to determine the subset $A \subseteq S$ that maximises the total value of the items in $A$ while the total weight of the items is less than a given limit $t$. 

CDF smoothing considers a key set $K$ of size $n$. We aim to find a subset $V$ (virtual points $k_{vi}$) of at most size $\lambda$ from a candidate set $C$ that would minimise the loss  $\mathcal{L}$ (which implies the maximisation of loss reduction from the loss without CDF smoothing). Naively, the set $C$ can be formed by considering $\lambda$ virtual point candidates between every two adjacent keys in $K$, i.e., $|C| \le \lambda \cdot (n-1)$.

To reduce the Knapsack problem to our CDF smoothing problem, the set $S$ of items is mapped to the candidate set $C$ of virtual points. We set the weight of every item (a candidate virtual point) to 1 and let $t$ be our target number of virtual points to be added, i.e., $\lambda$. The value of an item, $c_s$, is mapped to the loss reduction contributed by the corresponding virtual point. Maximising the values of the items in subset $A$ is mapped to maximising the total loss reduction of the virtual points in $V$. 
Due to the nature of our problem, the total loss reduction  when multiple virtual points are added together varies from the sum of the loss reduction when the virtual points are added individually. This can be represented as:
\begin{equation}
    |\sum_{s \in A} c_{s} |= 
    r\sum_{s \in A}|c_{s}|,
\end{equation}
where $|c_s|$ is the magnitude of the value of an item (a virtual point) and $r \in \mathbb{R}$ is a parameter. Here, $r$ is deterministic since it could be calculated based on Equation~\ref{eq:sse_ori}. As such, the total value of the combined items can be transformed to the sum of the individual item values, and maximising the latter for the Knapsack problem can be mapped to maximising the former for our problem.  

The transformation between the two problems can be done in polynomial time since there is a one-to-one mapping between them. Thus, when our problem is solved, the Knapsack problem is also solved in polynomial time. As such, our problem is NP-hard.
\end{proof}

Due to the computational complexity in finding an exact optimal solution for the learned index smoothing problem over a large set of keys, next, we consider two practical variants of the problem and propose highly effective heuristic solutions: 
(1) smoothing the CDF for the subset of keys $K_i$ indexed by an indexing function $f_i$ (Section~\ref{sec:smoothing_CDF}); and (2) smoothing the CDFs for all $m$ subsets $K_i$ when they are indexed under a hierarchical learned index (Section~\ref{sec:inv_pois_unbalanced}). 

\section{CDF Smoothing for a Single Linear Model}
\label{sec:smoothing_CDF}
We first consider a single indexing function over a segment of keys $K_i$. Given the smoothing budget $\lambda$, our optimisation goal is: 
\begin{equation}\label{eq:bilevel_opt}
\text{argmin}_{V_i, w,b}\mathcal{L}_{f_{w,b}}(K_i \cup V_i), \quad s.t. |V_{i}| \leq \lambda
\end{equation} 
This optimisation goal varies from Equation~\ref{eq:total_error}. Importantly, we now allow the slope ($w$) and the intercept ($b$) parameters of the indexing function $f$ to be refitted based on the keys $K_i$ (with adjusted ranks) and the inserted virtual points $V_i$, rather than just inserting  virtual points to adjust $rank(k)$ for $k \in K_i$ to fit $f(k)$ of the original indexing function $f$. This way, intuitively, we could achieve a lower loss with fewer virtual point insertions (hence reducing space overhead), as opposed to the naive optimal smoothing scheme described above that simply spreads $K_i$ to fit $rank(k)$ to a given $f(k)$.

In Equation~\ref{eq:bilevel_opt}, we include $V_i$ in the loss calculation, such that the storage space allocated to the virtual points can be used to accumulate data insertions, with minimized prediction errors when querying the inserted data points. 
 
\textbf{Challenges.} Allowing to refit the slope  and the intercept of the indexing function makes the optimisation problem even more difficult, as now the loss reduction brought by inserting a virtual point depends further on the other virtual points to be inserted. To solve the problem, a simple greedy heuristic is to iteratively select the best remaining candidate virtual point that leads to the largest loss reduction and to refit the indexing function after each virtual point selection.
This process is repeated for $\lambda$ iterations to identify all $\lambda$ virtual points to be inserted. 

There are complexity issues with this simple heuristic. 

\underline{Challenge 1: Large size of the candidate set.} Given keys $K_i$, the candidate virtual points can be any key value in $(\min\{K_i\}, \max\{K_i\})$ (detailed in Section~\ref{subsec:fvpc}), which can be a large range in real datasets. 

\underline{Challenge 2: Repeated loss calculations.} For each candidate virtual point $k_v$, we need to recalculate  $\mathcal{L}_{f}(K_i\cup V_i)$, where $V_i$ includes $k_v$, to help select the best candidate virtual point that leads to the largest loss reduction, which takes $O(|K_i|+\lambda)$ time. For $\lambda$ iterations and 
$O(p)$ candidate virtual points to be examined per iteration, overall,  the greedy solution above takes $O((|K_i|+\lambda) \cdot \lambda \cdot p)$ time.  

To overcome these challenges,  we present an efficient solution below to reduce the overall time complexity through (1)~reducing the number of candidate virtual points $p$ and (2)~the time cost to calculate the loss for each candidate virtual point.

\textbf{Our solution.} Our solution takes three steps: 

\textbf{Step 1:} We propose an effective approach to reduce the number of candidate virtual points. The idea is that, if there are consecutive candidate virtual points, we only need to consider the point among these consecutive points that minimises the loss function, which can be identified utilising the first-order partial derivative of the loss function (addressing Challenge 1, detailed in Section~\ref{subsec:fvpc}).

\textbf{Step 2:} We propose an efficient algorithm to calculate the loss. The core idea is to calculate the loss incrementally, and to reuse part of the calculation from the previous iteration, as only one new candidate virtual point is included into the calculation each time (addressing Challenge 2, detailed in Section~\ref{subsec:insert_one_point}).

\textbf{Step 3:} Finally, from the reduced set of candidate virtual points, and with efficient loss calculation, we present an efficient algorithm to compute the best subset of virtual points of size $\lambda$ (Section~\ref{subsec:ilvp}).

\begin{figure}[t]
	\centering
	
 \includegraphics[width=6cm]{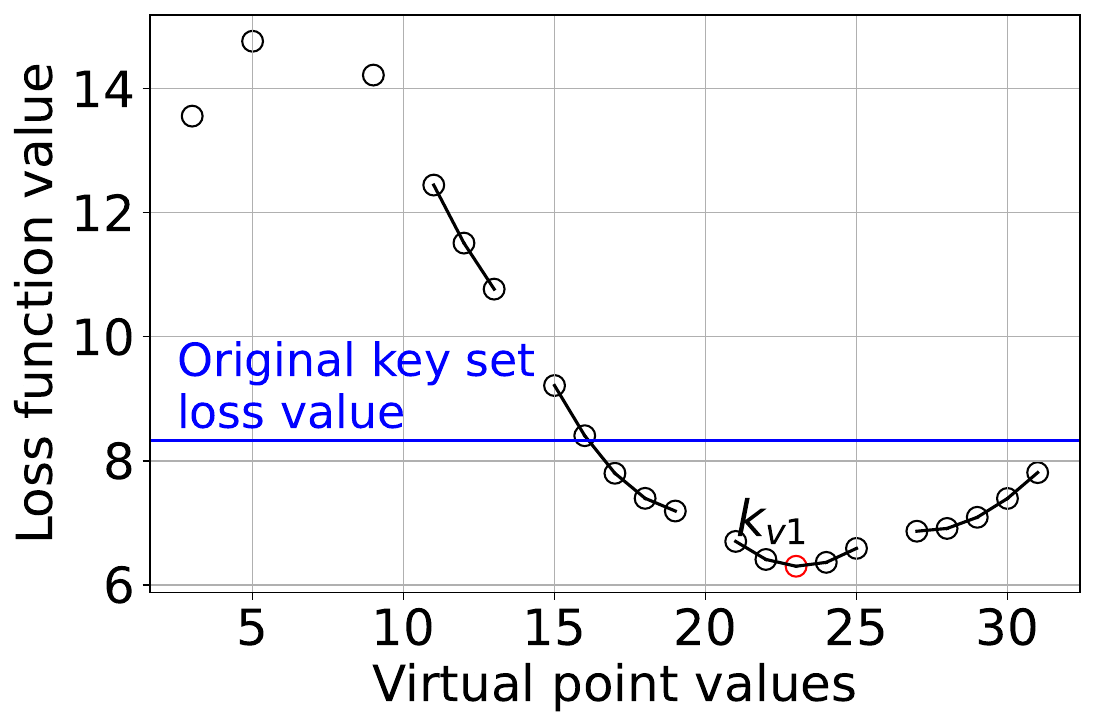}
 \caption{Loss function (SSE) value corresponding to different insertion positions for a virtual point.}
 \label{fig:loss_function}
\end{figure}

A running example on finding the best virtual point is shown in Fig.~\ref{fig:loss_function}, which corresponds to the keys in Fig.~\ref{fig:before_smooth}. In Fig.~\ref{fig:loss_function}, each hollow dot represents a candidate virtual point. Its $y$-value represents the loss (i.e., SSE) if the virtual point is included into the key set. Adjacent hollow dots are linked together, forming a segment of candidate virtual point values. For example, the segment formed by 21 to 25 is between index keys 20 and 26 (the index keys themselves are \emph{not} considered to be candidate virtual points; they correspond to the gaps between the curve segments in Fig.~\ref{fig:loss_function}). The goal is to search for the virtual point with the smallest loss value. In the given example, point 23 is the search target. 

In what follows, we present the algorithm to efficiently calculate the loss first (Section~\ref{subsec:insert_one_point}), based on which the strategy to reduce the candidate virtual points is described (Section~\ref{subsec:fvpc}), and our overall algorithm to compute the best $\lambda$ virtual points is detailed (Section~\ref{subsec:ilvp}). 

\subsection{Efficient Loss Calculation and Indexing Function Refitting} \label{subsec:insert_one_point}
We start with efficiently calculating the loss  for each candidate virtual point, where the idea is to reuse the calculations as much as possible. For the rest of Section~\ref{sec:smoothing_CDF}, we abuse the notation slightly and use $K$ instead of $K_i$ to denote the segment of keys for which CDF smoothing is to be done, since the discussion only concerns a single segment. Similarly, we use $f$ instead of $f_i$ to denote the indexing function, $\mathcal{L}$ instead of $\mathcal{L}_i$ to denote the loss function, and $V$ instead $V_i$ to denote the candidate virtual points for the segment. 

Let the rank of a key $k_i$ be $y_i$ (assuming that the ranks start from 0) for the $n$ keys in $K$, i.e., $y_i = rank(k_i)$. 
First, consider a virtual point $k_v$ with its rank (after inserted into $K$) being $y_v$ (we explain the case for multiple virtual points later). The loss for $K$ and $V = \{k_v\}$ can be calculated as follows, where $w$ and $b$ are the parameters of the indexing function $f$:
\begin{equation}
\label{eq:loss_ori}
    \mathcal{L}(\{K\cup V\}) = \sum_{i=1}^{n}(wk_{i}+b - y_i)^{2} +(wk_{v}+b - y_v)^{2}
\end{equation} 

It is computationally expensive to calculate the loss for each candidate virtual point using Equation~\ref{eq:loss_ori}, as $w$ and $b$ can change when the indexing function $f$ is refitted for the candidate virtual point, which is governed by the following equations (derived from the first-order partial derivatives of Equation~\ref{eq:loss_ori}):
\begin{equation}
    w = \frac{\sum_{i=1}^{n+1}(k_{i}-\bar{k_{v}})(y_{i}-\bar{y_{v}})}{\sum_{i=1}^{n+1}(k_{i}-\bar{k_{v}})^{2}},
    \label{eq:slope_ori}
\end{equation}

\begin{equation}
    b = \bar{y_{v}} - w\bar{k_{v}}.
     \label{eq:intercept_ori}
\end{equation} 
Here, $\bar{k_{v}}$ and $\bar{y_{v}}$ are the mean of the key set (i.e., $K\cup V$) and the rank set after inserting the virtual point $(k_v, y_v)$,  respectively. They can be computed by Equations~\ref{eq:mean_keyv} and~\ref{eq:mean_rankv} as follows:
\begin{equation}
    \bar{k_{v}} = \frac{\sum_{i=1}^{n}k_{i} + k_{v}}{n+1},
    \label{eq:mean_keyv}
\end{equation}

\begin{equation}
    \bar{y_{v}} = \frac{\sum_{i=1}^{n}y_{i} + n}{n+1}.
    \label{eq:mean_rankv}
\end{equation} 

Next, we rewrite Equation~\ref{eq:loss_ori} such that the candidate virtual point $k_v$ is separated from the values of $K$. This enables us to calculate the terms in the equation related to $K$ separately and then reuse their values for different candidate virtual points. This would reduce the time when computing the loss for different candidate virtual points.
\begin{equation}
\begin{split}
\label{eq:loss}
    \mathcal{L}(\{K \cup V\}) = w^{2}\sum_{i=1}^{n}k_{i}^{2} +2wbn\bar{k} \\ - 
    2w\sum_{i=1}^{n}k_{i}y_{i}+nb^{2} - 2nb\bar{y} \\ +\sum_{i=1}^{n}y_{original_i}^{2} + n^{2} - y_{v}^{2} + (w{k_v}+b-y_{v})^{2},
    \end{split}
\end{equation}
where
\begin{equation}
    \sum_{i=1}^{n}y_{i} = \sum_{i=1}^{n}y_{original_i}+n -y_{v},
    \label{eq:sum_rank}
\end{equation}

\begin{equation}
   \bar{y} = \frac{\sum_{i=1}^{n}y_{i}}{n},
\end{equation}

\begin{equation}
    \bar{k} = \frac{\sum_{i=1}^{n}k_{i}}{n},
\end{equation}

\begin{equation}
    \sum_{i=1}^{n}k_{i}y_{i} = \sum_{i=1}^{n}k_{i}y_{original_i} + \sum_{i=y_{v}}^{n}k_{i}.
    \label{eq:sum_keyrank}
\end{equation}
In the equations, $\bar{k}$ and $\bar{y}$ are the mean of the key set $K$ and the corresponding rank set before inserting the virtual point $(k_v, y_v)$, respectively; $y_{original_i}$ refers to the rank of key $k_i$ prior to inserting the virtual point.

Given the rewritten loss, Equations~\ref{eq:slope} and~\ref{eq:intercept} are derived by separating the terms related to the candidate virtual point from those related to the original key set $K$, which enable more efficient computation of $w$ and $b$.  
\begin{equation}
    w = \frac{(\sum_{i=1}^{n}k_{i}y_{i}+k_{v}y_{v}) - (n+1)\bar{k_{v}}\bar{y_{v}} }{(\sum_{i=1}^{n}k_{i}^{2}+k_{v}^{2})-(n+1)\bar{k_{v}^{2}}},
    \label{eq:slope}
\end{equation}

\begin{equation}
    b = \bar{y_{v}} - w\bar{k_{v}}.
     \label{eq:intercept}
\end{equation} 

\textbf{Adjustment for multiple virtual points.} In the case of inserting $\lambda$ virtual points, after inserting one virtual point $(k_{v1}, y_{v1})$, to find the next virtual point $(k_{v2}, y_{v2})$, the original key set terms will be adjusted to include the newly added virtual point $(k_{v1}, y_{v1})$. As a result, the loss for the next candidate virtual point $k_{v2}$ can also be efficiently calculated by considering only the changes induced by $k_{v2}$ and its corresponding $y_{v2}$.

In the derived equations above, the terms are separated based on whether they belong to the original key set or dependent on the candidate virtual key to be inserted. Doing so enables efficient calculations of the loss by reusing the terms of the original key set after calculating them just once.

\subsection{Filtering Virtual Point Candidates} \label{subsec:fvpc}

Next, we present an efficient approach to identify  the candidate virtual points that can potentially reduce the loss, thus providing 
a much smaller set of candidate virtual points to consider.  
This step is important because while using the equations above helps improve the efficiency of processing one candidate virtual point, the number of candidate virtual points to be processed has a multiplicative impact to the overall algorithm time efficiency. 

To reduce the search space for the candidate virtual points, we bound it in $(\min\{K\}, \max\{K\})$. This is because any virtual points added prior to $\min\{K\}$ would cause all keys' ranks to increase at the same time, while adding virtual points after $\max\{K\}$ would not impact any key's rank. As such, neither would help achieve a better-fitted indexing function. We also skip the index keys already in $K$, such that our solution can be compatible with learned indexes that do not support duplicate keys~\cite{sun2023learned}. 

\begin{figure}[t]
	\centering
	
 \includegraphics[width=6cm]{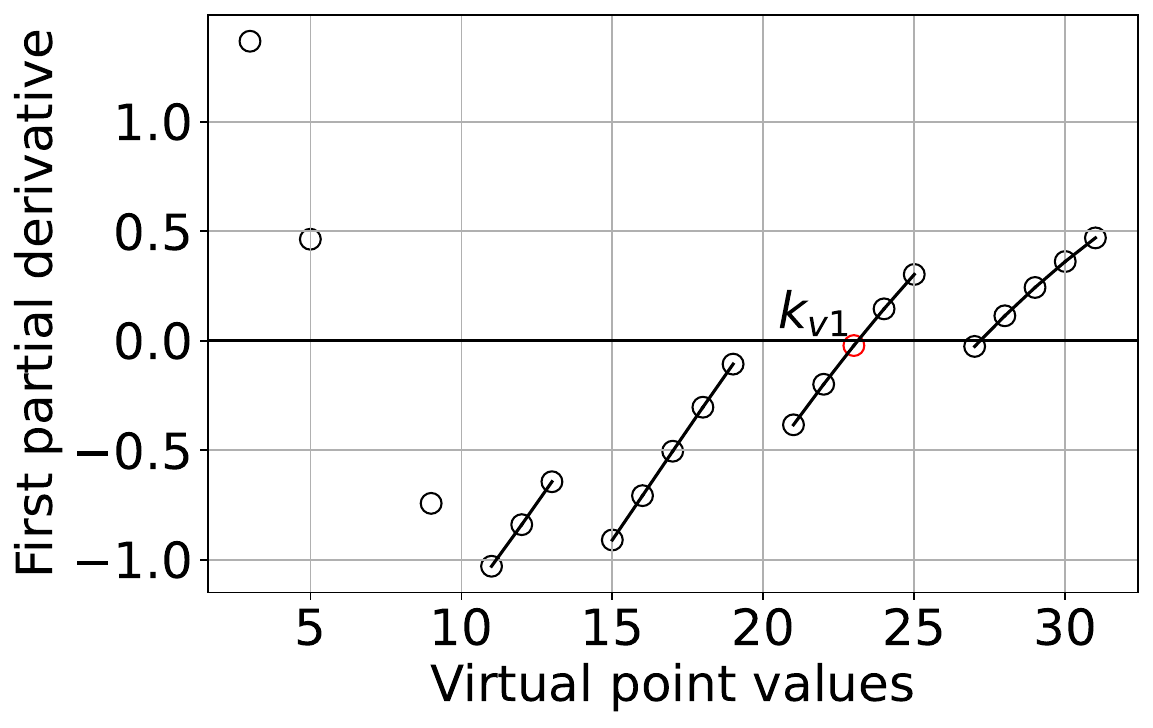}
 \caption{First-order partial derivatives of the loss (Equation~\ref{eq:loss_derv}) with respect to the key value of a virtual point $k_v$}
 \label{fig:loss_function_derv}
\end{figure}

Below, we present an approach based on the derivative of the loss function to further reduce the set of candidate virtual points. 
Our idea is illustrated using Fig.~\ref{fig:loss_function_derv}, which plots the partial derivative of the loss function with respect to a candidate virtual point $k_v$. Each sub-sequence (depicted as lines or dots) corresponds to the partial derivative of the loss of a sub-sequence of key values of a candidate virtual point (which can also be seen as a sub-sequence of candidate virtual points). 
The sub-sequences that cross the zero-value line (i.e., the $x$-axis)  imply that there is a minimal loss point within the sub-sequence. Otherwise, the minimal point for the sub-sequence must be at one of two endpoints of the sub-sequence, since each of such sub-sequences has been shown to be convex~\cite{kornaropoulos2022price}. Intuitively, this convex property is because the loss function is a summation of $n$ quadratic terms ($(wk_i + b - y_i)^2$) and only one non-quadratic term ($(wk_v + b - y_v)^2$), and the $wk_v$ term is nonlinear variable. We exploit this property to streamline the selection of candidate virtual points, i.e., to select the best virtual point from each sub-sequence. 

Following the idea above, we propose to 
filter the candidate virtual points as follows:
\begin{enumerate}
\item For each sub-sequence of candidate points (that is, where the candidate virtual point values are continuous), if the length of the sub-sequence is greater than 2, there can be a candidate virtual point within the sub-sequences
with a local minima of the loss. We compute the partial derivative of the loss function, which includes the previously added virtual points ($\mathcal{L}(\{K\cup V\})$ shown in Equation~\ref{eq:loss_ori}) with respect to candidate virtual point $k_v$, i.e., ${\mathcal{L}(\{K \cup V\})}'$ of the two endpoints of such sub-sequence (we present efficient ways to compute the partial derivative of a candidate later). 

\begin{itemize} 
\item If the sign of ${\mathcal{L}(\{K \cup V\})}'$ of the two endpoints are the same, it means that there is no point with the local minima within this sub-sequence (i.e.,, the sub-sequence does not cross the zero value in $y$-axis as shown in Fig.~\ref{fig:loss_function_derv}). In that case, we only need to consider the two endpoints of the sub-sequence as the candidate virtual points, and we can safely discard all the candidate virtual points in between. 

\item Otherwise, if the sign of ${\mathcal{L}(\{K \cup V\})}'$ of the two endpoints are opposite, it means that there is a point with a local minima within the sub-sequence. The minimal point can be calculated by using the two partial derivative values to find their intersection with the x-axis. As this minimal point is guaranteed to have a smaller loss than all the other points in that sub-sequence, only the point is considered as a candidate virtual point. All other points in the sub-sequence can be safely discarded.
\end{itemize}

\item If the length of the sub-sequence is less than or equal to 2, we need to consider all points in the sub-sequence as the candidate virtual points.
\end{enumerate}

\textbf{Efficient computation of the first-order derivative of the loss function value.} Similar to the computation of the loss function, we present an efficient way to calculate the first-order partial derivative of the loss function with respect to candidate virtual point $k_v$, where  $k_v$ and its related terms are separated from the other terms to enable reusing the terms that require information from $K$ only. 
This is achieved by the following equation: 
\begin{equation}
\begin{split}
    \label{eq:loss_derv}
    {\mathcal{L}(\{K \cup V\})}' = 2({w}'(w\sum_{i=1}^{n}k_{i}^{2}+nb\bar{k}-\sum_{i=1}^{n}k_{i}y_{i}) + \\ 
    n{b}'(w\bar{k} + b -\bar{y}) +(wk_{v}+b-y_{v})({w}'k_{v}+w +{b}')).
    \end{split}
\end{equation}
Here, $w'$ and $b'$ refer to the partial derivatives of $w$ and $b$ with respect to $k_v$, respectively. They can be computed by Equations~\ref{eq:slope_derv} and~\ref{eq:inter_derv} as follows.
\begin{equation}
    {w}' = \frac{A(n(y_{v}-\bar{y})) - B (2n(k_{v}-\bar{k}))}{A^{2}},
    \label{eq:slope_derv}
\end{equation}

\begin{equation}
    {b}' = -\frac{(w+(n+1)\bar{k_{v}}{w}')}{n+1}
    \label{eq:inter_derv},
\end{equation}

\begin{equation}
   A = (n+1)(\sum_{i=1}^{n}k_{i}^{2}+k_{v}^{2}) - ((n+1)\bar{k_{v}})^{2},
\end{equation}

\begin{equation}
    B = (n+1)(\sum_{i=1}^{n}k_{i}y_{i}+k_{v}y_{v}) - (n+1)^{2}\bar{k_{v}}\bar{y_{v}}. 
\end{equation}
Here, $A$ and $B$ are intermediary for computing the partial derivatives.

\subsection{Algorithm for Inserting $\lambda$ Virtual Points} \label{subsec:ilvp} 
After filtering the candidate virtual points, among the remaining ones, 
we present an efficient algorithm to find the best subset of candidate virtual points of size $\lambda$.

When there are $\lambda$ virtual points to insert, the optimal solution would require computing the loss for every size-$\lambda$ subset of the candidate virtual points in the range of $(\min\{K\}, \max\{K\})$. If there are $p$ possible insertion positions for the virtual points, the time complexity will be $O(^pC_\lambda \cdot n \cdot p)$, where $^pC_\lambda$ is the combination of every size-$\lambda$ subset from $p$. As this will be prohibitively expensive for a large dataset, we propose a greedy algorithm that inserts individual virtual points iteratively.  

Our core idea is to identify the virtual point that would minimise the  loss for each sub-sequence (i.e., local minima for the sub-sequences) and select the one that reduces the loss the most (i.e., global minimum). This process needs to be performed $\lambda$ times. The algorithm for CDF smoothing by inserting $\lambda$ virtual points is summarised in Algorithm~\ref{alg:poi} and described below. 

\begin{algorithm}
\begin{small}
\caption{CDF\_smoothing}\label{alg:poi}
\begin{algorithmic}[1]
\Require Key set: $K$, loss function with new virtual point: $\mathcal{L}( K \cup k_v)$, smoothing threshold: $\alpha$
\State $U$, $C$, $V = [ ]$
\State $G = [ ]$, $M = [ ]$  \Comment{Arrays of point pairs}
\State $\mathcal{L}'(K \cup k_v) = \frac{\partial L(K \cup k_v)}{\partial k_v}$, $\lambda = \alpha \cdot K.size $, $\mathcal{L}_{previous} = \mathcal{L}(K \cup \varnothing) $
\State Find the endpoint pairs, $E$, for each sub-sequence

\While{$V.size < \lambda$}
\For{$i$ from 1 to $E.size$} \Comment{Separate sub-sequences}
\If{$E[i].second$ - $E[i].first \le 1$}
    \State Append $E[i].first$ and $E[i].second$ to $C$
\Else \Comment{There are more than 2 points}
   \State Append $E[i].first$ to $G.first$ and $E[i].second$ to $G.second$

\EndIf
\EndFor

\For{$i$ from 1 to $G.size$} \Comment{Calculate the partial derivatives}
\If{$\mathcal{L}'(K \cup G[i].first) \cdot \mathcal{L}'(K \cup G[i].second) <$ 0} 
    \State Append $G[i].first$ to $M.first$ and $G[i].second$ to $M.second$
\Else
     \State Append $G[i].first$ and $G[i].second$ to $C$
\EndIf
\EndFor

\For{$i$ from 1 to $M.size$} \Comment{Calculate minimum point}
     \State Append $minimum\_point(M[i].first$, $M[i].second)$ to $C$
\EndFor

\For{$i$ from 1 to $C.size$} \Comment{Calculate loss value}
     \State $U[i]$ = $\mathcal{L}(K \cup C[i])$
\EndFor

\State Find index $i$ of minimum $\mathcal{L}$
\If{$\mathcal{L}_{previous} \le U[i]$} 
    \State break
\EndIf
\State Append $C[i]$ to $V$, Append $C[i]$ to $K$, $\mathcal{L}_{previous} = U[i]$
\EndWhile

\State  \Return C
\end{algorithmic}
\end{small}
\end{algorithm}

\textbf{The algorithm.} Our algorithm takes as input a key set $K$ and a smoothing threshold $\alpha$ (or a smoothing budget $\lambda = \alpha n$). We use $G$ to denote a set that stores the potential sub-sequences, where there can be a candidate virtual point within the sub-sequences with a local minima of the loss. The candidate virtual points contributing the local minima are stored in an array $M$, while $C$ stores the set of candidate keys for the virtual points. We use 
$U$ to hold the loss for each candidate virtual point and vector $V$ to store the final optimal virtual points.

First, the algorithm identifies the sub-sequences of candidate virtual points that could have their minimal loss at the endpoints or in-between the sequence. This is shown in Lines 4 to 12. If there are more than two points in a sub-sequence, the candidate virtual point with the minimal  loss can be within that sub-sequence. As such, the two endpoints of the sub-sequence are saved in array $G$ for calculating the partial derivatives. 

Afterwards, in Lines 13 to 22, the partial derivative of the loss function  with respect to the candidate virtual points is calculated for all point pairs in $G$ using the equations derived above. If the signs of the partial derivatives  corresponding to the two endpoints of a sub-sequence are different (i.e., on opposite sides of the $x$-axis), the two endpoints are added to array $M$ for calculating the minimum point. As shown in Fig.~\ref{fig:loss_function_derv}, for the sub-sequences that contain candidate virtual points with minimal loss, the partial derivatives of the end points will appear on the two sides of the $x$-axis.
These minimal points are added to array $C$ after they are calculated. If the partial derivatives of the two endpoints are on the same side of the $x$-axis, the minima is at one of the endpoints, as such the two end points are added to $C$.

Finally, Lines 23 to 31 compute the loss for each candidate virtual point in $C$ and select the point with the minimum loss, as long as the new  loss is smaller than the existing loss obtained so far over $K$ and any previously inserted virtual points. 
This process is repeated until at most $\lambda$ virtual points are inserted, or when the loss is not reduced any further. When the algorithm terminates, the final virtual points in $V$ are returned.

\textbf{Complexity analysis.} Our proposed CDF smoothing algorithm reduces the computation of the loss over $K$ to just once, which takes $O(n)$ time. This process is repeated to find $\lambda$ optimal candidate virtual points. However, there is no need to recalculate the loss function after adding a virtual point, as we could treat the key set with the previous virtual point inserted as the new original or base key set for a constant time calculation. Thereby, giving a time complexity of $O(\lambda + n)$.

\section{CDF Smoothing for Hierarchical Indexes} \label{sec:inv_pois_unbalanced}

In this section, we present the CDF smoothing to a hierarchical learned index to improve the performance of queried keys. A direct application to individual nodes would help reduce leaf-node search time by improving the learnability of the models but fail to address traversal time. Therefore, a method for addressing both traversal and leaf-node search is required. As such we present \emph{\algo} to smooth segments of the CDF for different sub-trees in the hierarchical structure of a learned index in order to merge and reduce the overall structure height. A major challenge is the balancing between the improvement of traversal time due to the reduction of the index height and the increase in leaf-node search time due to more keys being merged into single nodes. To address this, we present a cost model that takes both of these factors into consideration. 

The core idea is to start from the bottom most level of the index that contains parent nodes of leaf-nodes and select those nodes. Then for each of these parents nodes, the keys in the node and its child nodes are collected, which are then subjected to smoothing using Algorithm~\ref{alg:poi}. If the minimum cost threshold is satisfied (more details below), then the sub-tree and the node are reconstructed to merge the collected nodes. The merging is performed by creating a new leaf-node in place of the parent node and placing the keys from the collected nodes. By doing so, more keys would be placed in upper level nodes of the index as the indexing functions of these nodes would be improved by the CDF smoothing, but the cost models would limit the number of keys as to not offset the performance gain by the increase in the leaf-node search time. Further details regarding this is given in Section~\ref{sec:cost_condition}. This process is performed until the root node depicted as level 1 is reached, thus reducing the loss  ($\mathcal{L}$). 

This process is applied to a constructed learned index structure as the purpose of the method is to enhance the structure of the index. Further, it is computationally expensive to perform the smoothing operation on the full key set. As such it is more reasonable to handle subsets of the key set in the constructed hierarchical index. For this purpose, unbalanced learned index structures are better suited as it gives the ability to reduce the height of taller branches without affecting the rest.

\begin{algorithm}[t]
\begin{small}
\caption{\algo}\label{alg:lipp_poi}
\begin{algorithmic}[1]
\Require Nodes with sub trees : $Nodes$, smoothing threshold : $\alpha$, cost threshold : $c$
\State $Nodes = [ ]$
\State $keyset = [ ]$
\State $keyset\_smooth = [ ]$

\State $max\_level$ $\gets$ maximum level of index with sub trees 
\State $current\_level$ $\gets$ $max\_level$
\While{$current\_level > 1$}
\State $Nodes \gets$ all nodes with sub trees

\For{$i$ from 1 to $Nodes.size$}
\State $keyset \gets$ collect all keys in the node and its sub tree
\State $keyset\_smooth \gets$ $CDF\_smoothing(keyset,\alpha)$ \Comment{Using Algorithm~\ref{alg:poi}}

\If{$cost < c$}  
\State Reconstruct the sub-tree and node with $keyset\_smooth$
\EndIf
\EndFor
\State $current\_level \leftarrow current\_level - 1$ 
\EndWhile
\end{algorithmic}
\end{small}
\end{algorithm}

The algorithm for CDF smoothing of a hierarchical learned index structure is given in Algorithm \ref{alg:lipp_poi} and described below. First the algorithm starting from the maximum level of the index, and identifies all nodes with sub-trees, which is shown in Lines 5-7. Afterwards, as shown in Lines 8-14, each of the node's and its sub-tree's keys are collected and subjected to the CDF smoothing. Provided that they meet the minimum cost threshold selected, the node and its sub-tree are reconstructed to promote as many keys to upper levels as possible. This process is iteratively performed in a bottom up manner for other levels of the index. 

\subsection{Cost Conditions}\label{sec:cost_condition}
For indexes that does not contain any searching component such as LIPP and SALI, their loss function values can be taken as the cost conditions. This is because if the new model could hold more keys than before, then it does not have any other component (that is, leaf-node search time) that would negatively affect the performance. However, for the indexes with leaf-node search components like ALEX, there must be a trade-off between the increase of leaf-node search time over the reduction of traversal time. The reason is, introducing new keys into the node would require more time to locate the key. 
For this purpose, we develop the following cost model, where reconstruction is performed only if the cost is less than a specified threshold value, $c$. 

\begin{equation}
\begin{split}
    cost = search\_constant \times expected\_number\_of\_searches \\ 
    + traversal\_constant \times index\_level
\end{split}
\end{equation}

To make the implementation hardware independent, the constants can be measured by sampling queries to measure the time spent per leaf-node search for the case of $search\_constant$ and the traversal time spent per level for $traversal\_constant$. The $expected \\\_number\_of\_searches$ can be calculated via the inbuilt function in ALEX that uses the $log_2$ error to estimate it. Considering the cost model depicts the expected query time for the node, the cost threshold, $c$ should be set below 0 to identify an improvement. Setting a lower value would result in fewer keys being able to be promoted to upper levels but the expected query time improvement will be greater.

\textbf{Complexity analysis.} For a key set of size $n$, with a smoothing budget of $\lambda$ and an index structure with $m$ non-leaf nodes, the complexity for the developed algorithm can be calculated as follows. 
The complexity for $node_i$ with $n_i$ keys and a smoothing budget of $\lambda_i$ is $O(\lambda_i + n_i)$. Similarly, for the $m$ nodes, we would get a complexity of $O(\lambda_1 + n_1 + \lambda_2 + n_2 + \cdots + \lambda_m + n_m)$. This can be simplified to $O(\lambda + n)$.

\textbf{Choice of smoothing threshold.} Increasing the smoothing threshold would make the algorithm insert more virtual points, thus reducing the loss function value of the newly fitted model even more. As a result these indexing functions would be able to accommodate more keys which were originally in lower levels of the index and improve their query time. However, that is a trade-off between the query time improvement and the higher space cost for increasing this threshold. Further, key sets that cause the original index structure to construct poorly should benefit more from a higher smoothing threshold as shown in the Experimental results~\ref{sec:smooth_impact}.

\subsection{Approximation Quality Analysis} 
\label{sec:analysis}

The effectiveness of our proposed greedy method of iteratively identifying the $\lambda$ virtual points as opposed to the exhaustive manner of comparing all $\lambda$ subsets, is demonstrated via experimentation in this section.

The key set of 10 keys given in Fig.~\ref{fig:scatter_keys} was subjected to CDF smoothing with a smoothing threshold ($\alpha$) of 0.5 (smoothing budget of 5) via both methods. The results are shown in Table~\ref{tab:app_quality}. Here, the greedy method improves the loss by 72.34\%, while the exhaustive method improves it by 74.44\%. However, the time taken by the exhaustive method is nearly 3 orders of magnitude more than the greedy method. This results show that the effectiveness of the greedy method is similar to the exhaustive method, and the exhaustive method is impractical to use in real datasets.

\begin{table}[b]
\caption{Approximation Quality Results}
    \label{tab:app_quality}
    \centering
\begin{tabular}{lrrr}
\toprule
           & Exhaustive  & \algo\  & Original \\  \midrule
Loss  & 2.118       & 2.293   & 8.327    \\ \hline
Time (ns)  & 140,656,167 & 424,667 & N/A \\ \bottomrule      
\end{tabular}
\end{table}

\section{Experimental results}
\label{sec:experiments}
Next, we report experimental results. The implementation of our evaluation framework is based on an existing benchmark~\cite{bach2022testingrobust} implementation. All experiments were run on an Ubuntu 20.04.5 virtual machine with an AMD EPYC 7763 64-Core CPU and 128 GB RAM.

\subsection{Experimental Settings}

\textbf{Competitors.} To show the general applicability of our proposed techniques, we integrate \textbf{\algo} with recent learned indexes, including \textbf{ALEX}~\cite{ding2020alex}, \textbf{LIPP}~\cite{wu2021lipp}, and \textbf{SALI}~\cite{ge2023sali} {(SOTA)}.
These three indexes were chosen because they are the latest and among the most widely used benchmark learned indexes. They have reported strong empirical performance, outperforming both traditional indexes such as the B$^+$-tree and learned ones~\cite{sun2023learned} such as the PGM index~\cite{ferragina2020pgm}, XIndex~\cite{tang2020xindex}, and FINEdex~\cite{li2021finedex}. For simplicity, we do not repeat the comparison results with these other indexes.

\textbf{Datasets.} We run experiments with four datasets from two benchmark works~\cite{kipf2019sosd,wongkham2022updatable}: 
(1)~\textbf{Facebook} contains 200 million integer Facebook user IDs~\cite{van2019fb_source}; (2)~\textbf{Covid} contains 200 million integer tweet IDs randomly sampled from tweets tagged with ``Covid-19''~\cite{lopez2021covid};
(3)~\textbf{OSM} contains 200 million locations randomly sampled from OpenStreetMap and represented as Google S2~\cite{s2_google} cell IDs~\cite{pandey2018osm_cell}; and 
(4)~\textbf{Genome} contains 200 million entries of loci pairs in human chromosomes represented as integers~\cite{rao2014genome}.
For all datasets, duplicate keys were removed to suit LIPP and SALI's requirements.

\begin{figure}[ht]
  \centering
  
  \begin{subfigure}[b]{0.11\textwidth}
    \includegraphics[width=\textwidth]{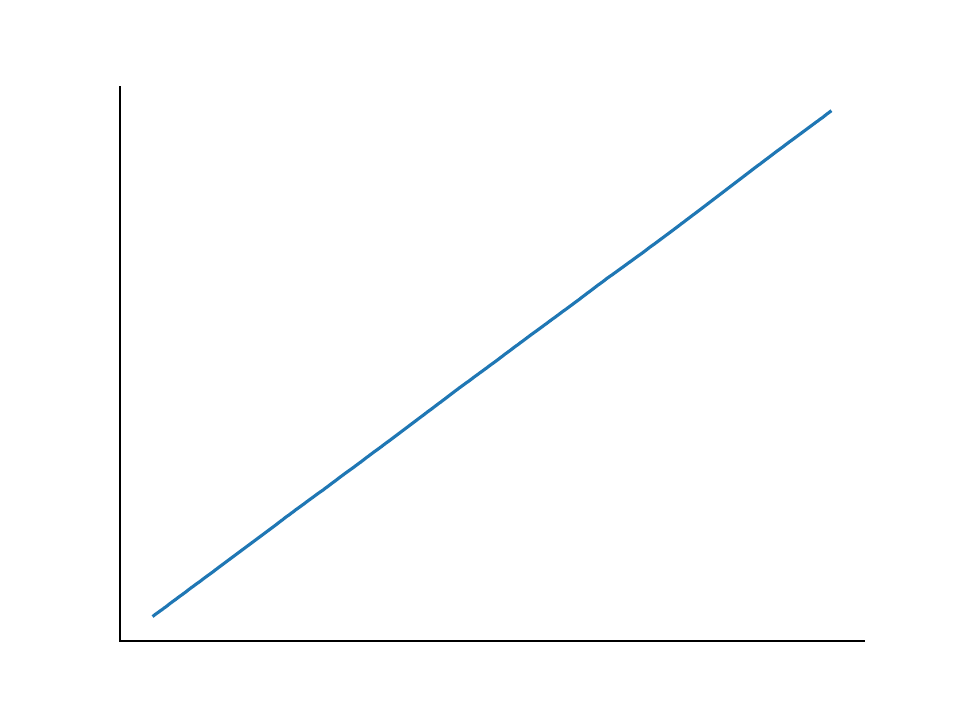}
    \caption{Facebook}
 \label{fig:cdf_fb}
  \end{subfigure}
  \begin{subfigure}[b]{0.11\textwidth}
    \includegraphics[width=\textwidth]{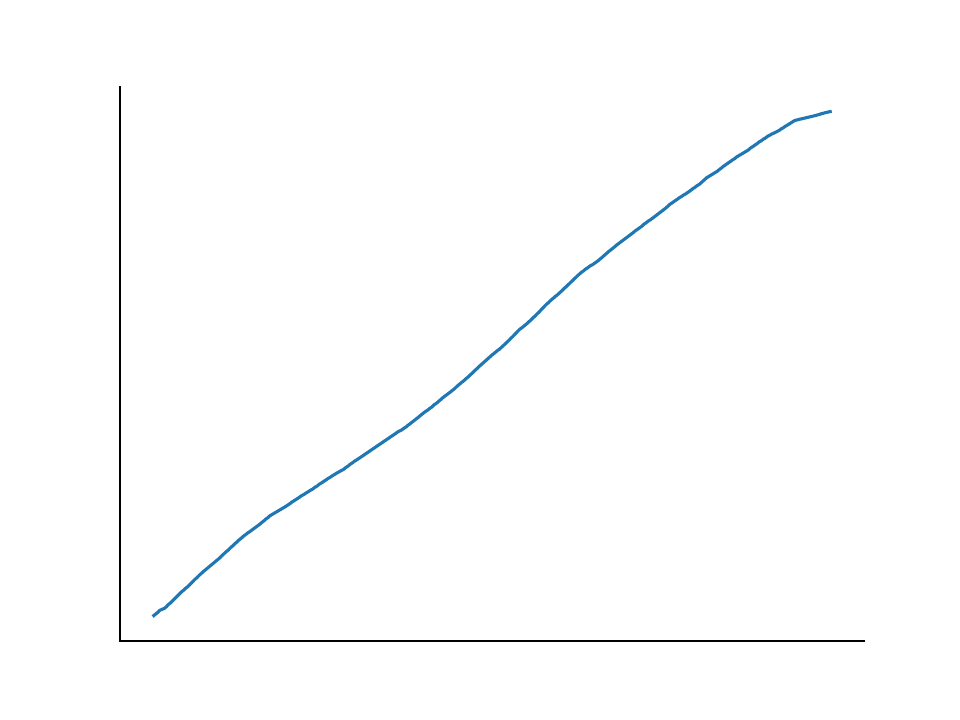}
    \caption{Covid}
 \label{fig:cdf_covid}
  \end{subfigure}
  \begin{subfigure}[b]{0.11\textwidth}
    \includegraphics[width=\textwidth]{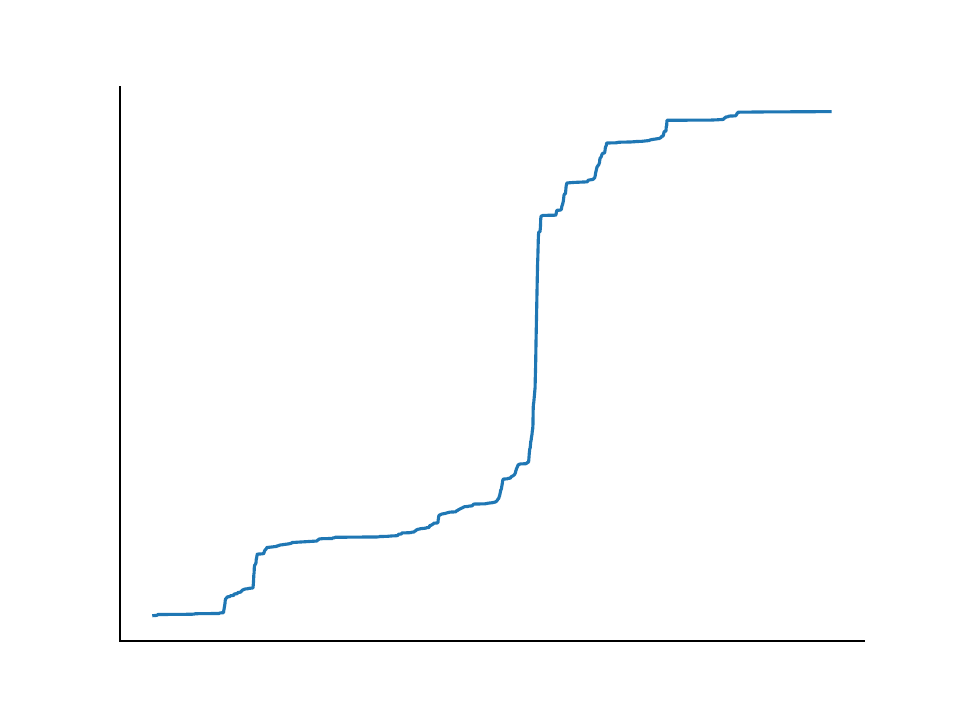}
    \caption{OSM}
 \label{fig:cdf_osm}
  \end{subfigure}
  \begin{subfigure}[b]{0.11\textwidth}
    \includegraphics[width=\textwidth]{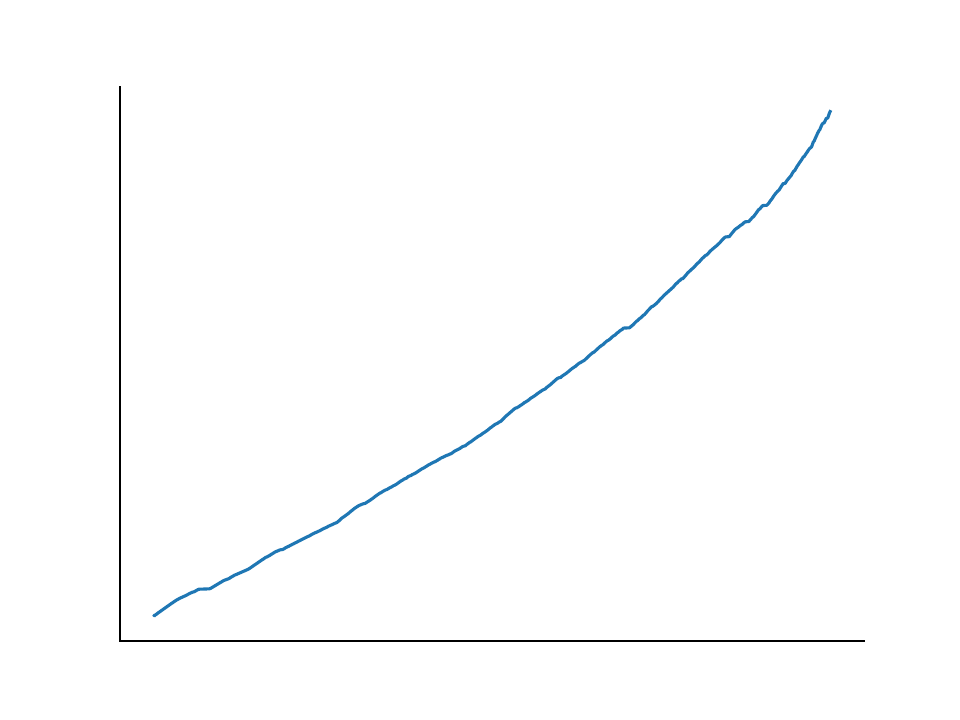}
    \caption{Genome}
 \label{fig:cdf_genome}
  \end{subfigure}
\\
  \begin{subfigure}[b]{0.11\textwidth}
    \includegraphics[width=\textwidth]{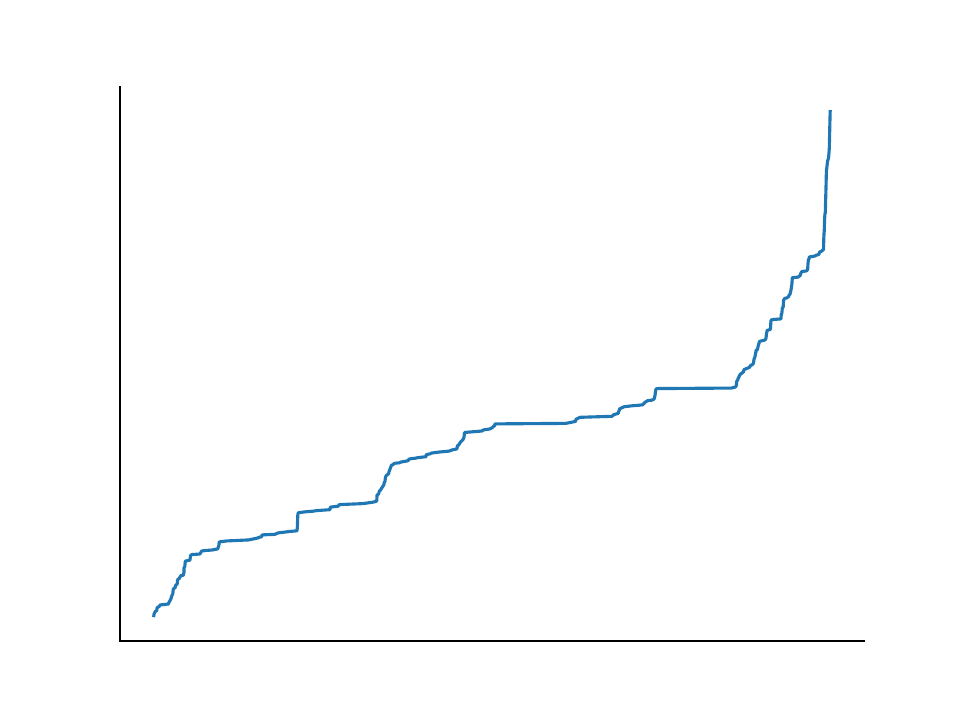}
    \caption{Facebook \\(zoomed-in)}
 \label{fig:cdf_fb_mid}
  \end{subfigure}
  \begin{subfigure}[b]{0.11\textwidth}
    \includegraphics[width=\textwidth]{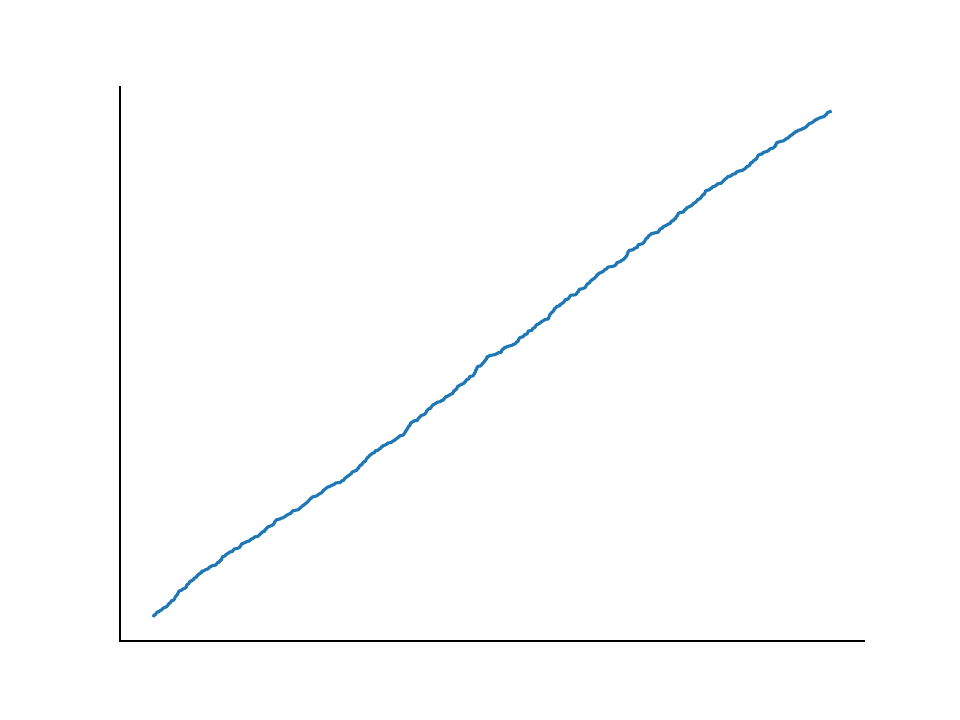}
    \caption{Covid \\(zoomed-in)}
 \label{fig:cdf_covid_mid}
  \end{subfigure}
  \begin{subfigure}[b]{0.11\textwidth}
    \includegraphics[width=\textwidth]{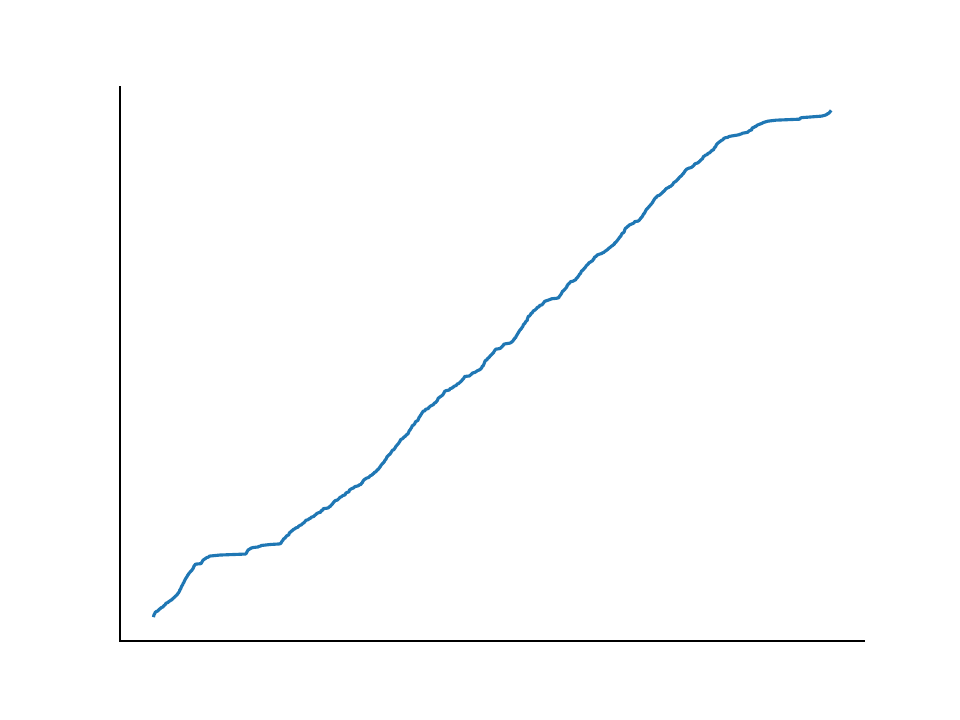}
    \caption{OSM \\(zoomed-in)}
 \label{fig:cdf_osm_mid}
  \end{subfigure}
  \begin{subfigure}[b]{0.11\textwidth}
    \includegraphics[width=\textwidth]{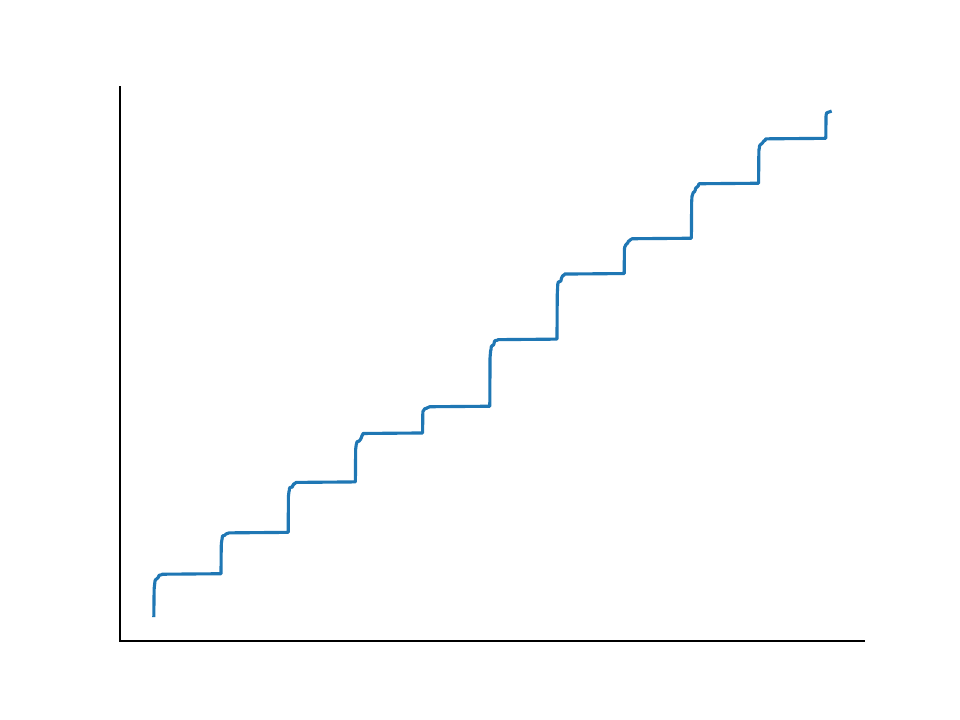}
    \caption{Genome \\(zoomed-in)}
 \label{fig:cdf_genome_mid}
  \end{subfigure}
  \caption{CDFs of the datasets}
 \label{fig:cdf_datasets}
\end{figure}

Out of the four datasets, OSM and Genome are considered more difficult for learned indexes~\cite{wongkham2022updatable} (hard datasets), while Facebook and Covid are easier (easy datasets). To illustrate this, the CDFs of the full datasets are plotted in Figs.~\ref{fig:cdf_fb} to~\ref{fig:cdf_genome}. All datasets except OSM have almost globally linear CDFs. Zooming in the CDFs shows that there is more variability in the local distribution patterns, as shown in Figs.~\ref{fig:cdf_fb_mid} to~\ref{fig:cdf_genome_mid} (each showing from  the 100 million-th data point to the next thousand data points). Except for Covid, all datasets deviate from linear CDFs at local level, especially Genome. 

\textbf{Workloads.} 
We use the following two types of workloads: 

(1)~\emph{Read-only workload.} 
The learned indexes ALEX, LIPP, and SALI are constructed over the full datasets. Afterwards, our \algo\ algorithm is applied to optimise their structures. Then, the queries (detailed below) are run. 

(2)~\emph{Read-write workload.} 
The learned indexes are constructed over a random half of each dataset and then \algo\ is applied. The other half of the dataset is inserted in random batches of size $0.1n$. Queries are run after each batch insertion without using \algo\ again for each batch. 

\textbf{Queries.} Considering the main objective of the developed method is to improve the performance of keys in lower levels, the experiments are focused on them. Specifically, we report results for the \textbf{promoted data}, which includes every key that has been promoted to upper levels in the index by our algorithm.

\textbf{Parameters.}
We vary the smoothing threshold, $\alpha$,  from 0.05 to 0.8, with a default value of 0.1. To show the scalability of our algorithm by varying the dataset size, the original datasets were down sampled by eliminating every $j$-th key from the sorted datasets in order to remove $n/j$ data points and create smaller datasets of size 12.5 million, 25 million, 50 million, and 100 million, respectively. The default datasets are the original ones with 200 million points.

For each queried key, the query time was recorded by repeating the query 100 times and taking its average. 

For LIPP and SALI, they can create nodes that are indexing only a few keys~\cite{wu2021lipp}. For these two indexes, \algo\ is run starting at the second level of the index structures, such that each smoothing step can benefit more points. This is not an issue for ALEX, and \algo\ is run starting at the bottom level. Further, since the query times of the keys in the top two levels of the index structures are very close, \algo\ stops at the second level from the top (i.e., the root).

  \begin{figure*}[t]
  \centering
  \captionsetup[subfigure]{justification=centering}
  \begin{subfigure}[b]{0.35\textwidth}
    \includegraphics[width=\textwidth]{figs/legend.pdf}
  \end{subfigure} \\
 \begin{subfigure}[b]{0.32\textwidth}
    \includegraphics[width=\textwidth]{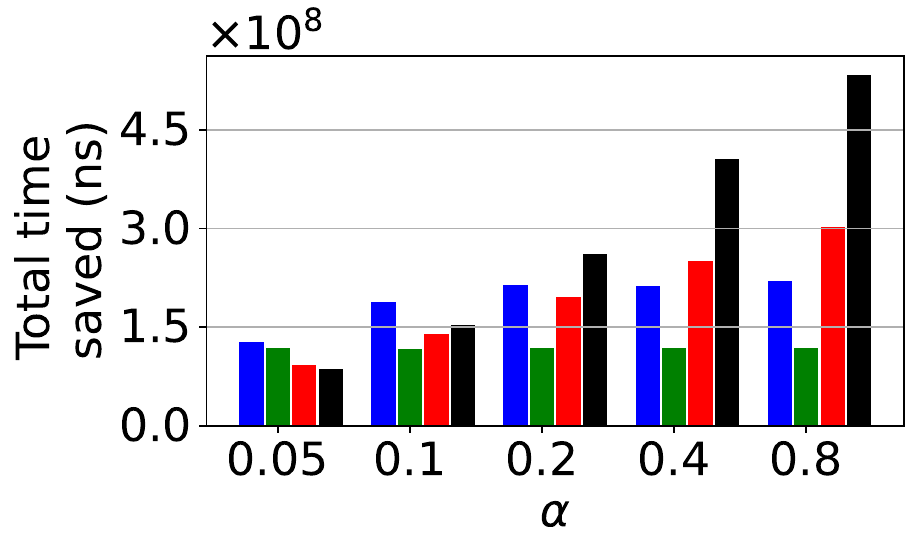}
    \caption{LIPP}
 \label{fig:query_improvement_lipp}
  \end{subfigure}
  \hfill
  \begin{subfigure}[b]{0.32\textwidth}
    \includegraphics[width=\textwidth]{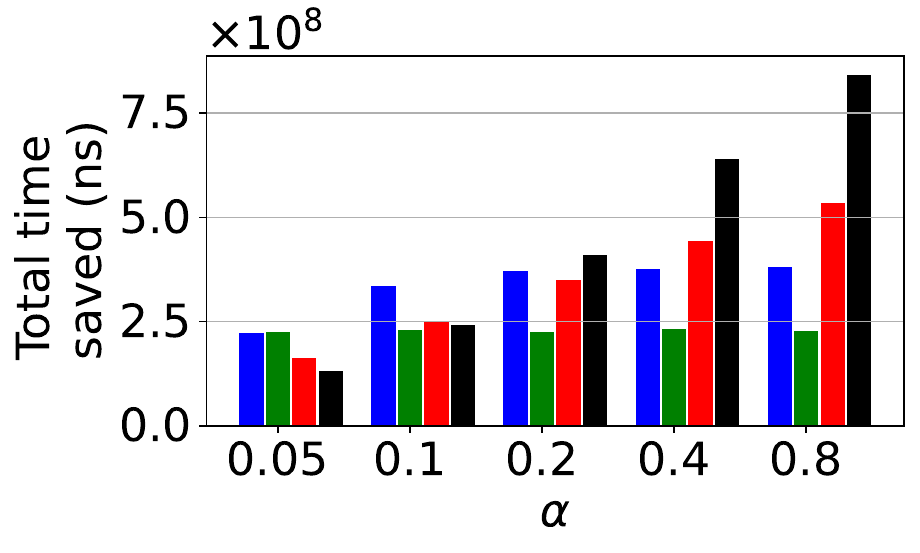}
    \caption{SALI}
 \label{fig:query_improvement_sali}
  \end{subfigure}
  \hfill
  \begin{subfigure}[b]{0.32\textwidth}
    \includegraphics[width=\textwidth]{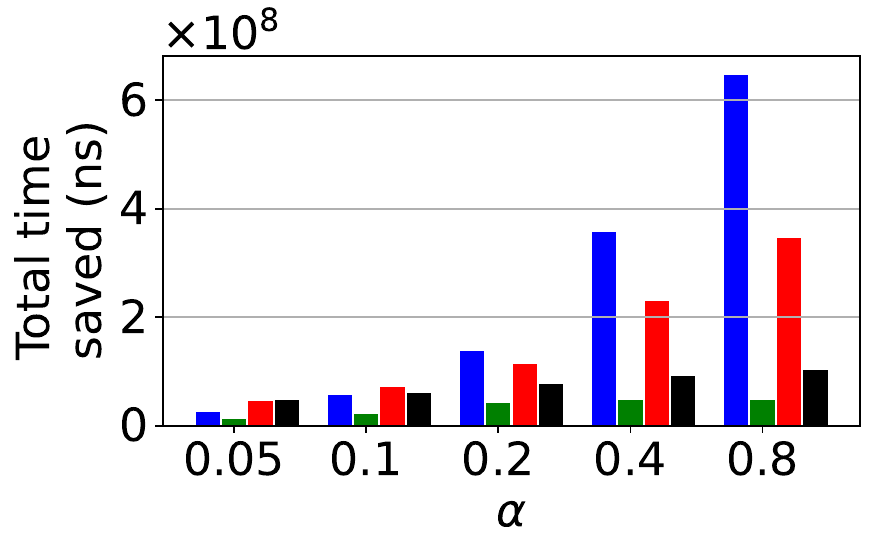}
    \caption{ALEX}
 \label{fig:query_improvement_alex}
  \end{subfigure}
  \caption{Total time saved vs. smoothing threshold $\alpha$}
 \label{fig:total_saved}
\end{figure*}

\begin{figure*}[t]
  \centering
  
  \captionsetup[subfigure]{justification=centering}
 \begin{subfigure}[b]{0.32\textwidth}
    \includegraphics[width=\textwidth]{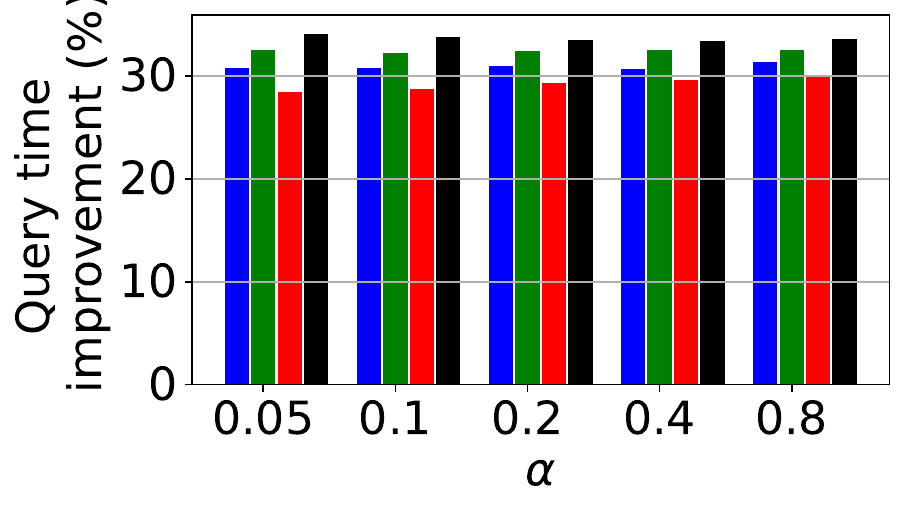}
    \caption{LIPP}
 \label{fig:query_difference_lipp}
  \end{subfigure}
  \hfill
  \begin{subfigure}[b]{0.32\textwidth}
    \includegraphics[width=\textwidth]{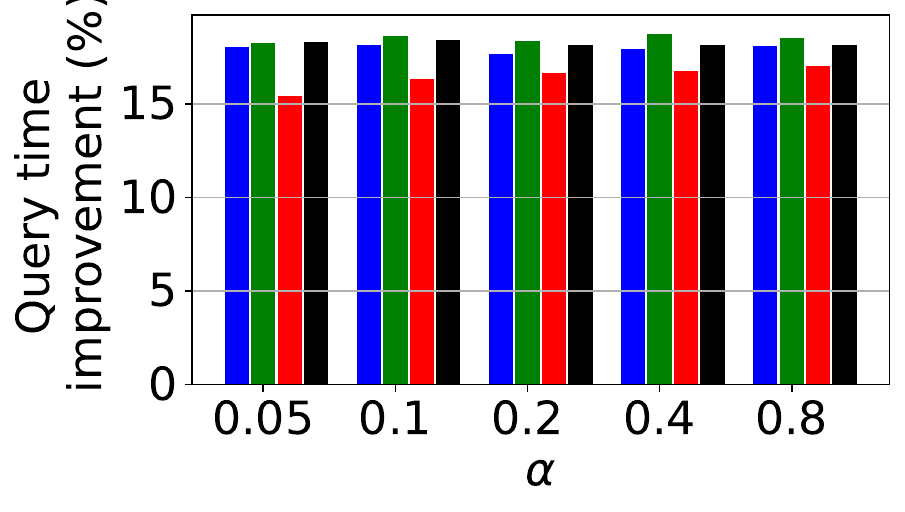}
    \caption{SALI}
 \label{fig:query_difference_sali}
  \end{subfigure}
  \hfill
  \begin{subfigure}[b]{0.32\textwidth}
    \includegraphics[width=\textwidth]{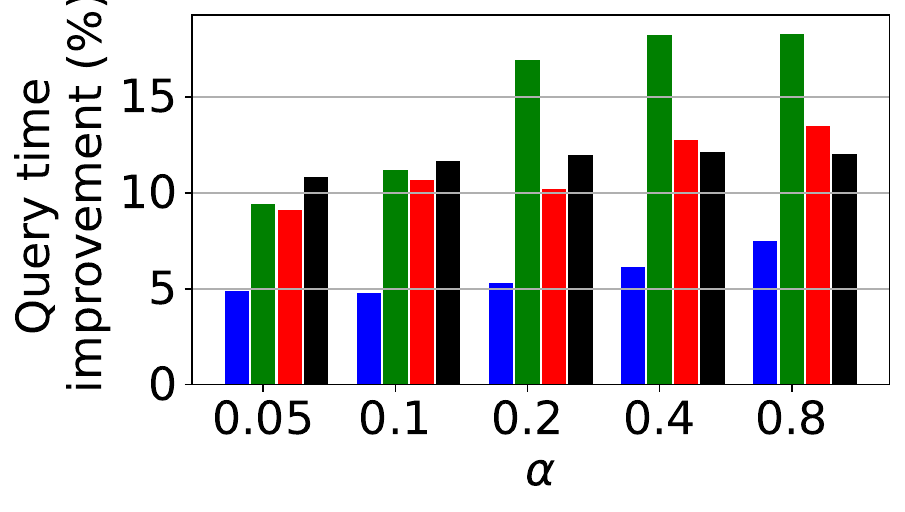}
    \caption{ALEX}
 \label{fig:query_difference_alex}
  \end{subfigure}
  \caption{Query time improvement vs. smoothing threshold $\alpha$}
 \label{fig:relative}
\end{figure*}

\textbf{Evaluation metrics.} We report: (1)~the \textbf{total query times saved}  by the \algo-enhanced indexed compared with those of the original indexes; (2)~the \textbf{query time improvement~(\%)}, which is the relative difference between the average query time (over all queried keys) of the \algo-enhanced indexes and that of the original ones; (3)~the \textbf{promoted data~(\%)}, which is the percentage of keys promoted to upper levels in the index structure among all keys that can be promoted (i.e., keys at levels 3 or below of the original indexes); (4)~the \textbf{storage space increase (\%)}, which is the relative index size overhead of the \algo-enhanced indexes compared with the original ones; (5)~the \textbf{node reduction (\%)}, which is the percentage of nodes reduced by the \algo-enhanced indexes over the original ones; and (6)~the \textbf{insert time increase (\%)}, which is relative increase in the average time per insertion required by the \algo-enhanced indexes compared to the original ones.

\subsection{Results on Read-only Workloads}
\label{sec:runtime}

\subsubsection{Impact of Smoothing Threshold} 
\label{sec:smooth_impact}
We vary the smoothing threshold from 0.05 to 0.8 to quantify its impact.

\textbf{Query time (for promoted data).} 
Here, we report the query time improvement by CSV for the `promoted keys' (i.e., the keys that is promoted to an upper level of the index by CSV), compared to the original index. 
We depict the total time saved due to the method in Fig.~\ref{fig:total_saved}. The general trend is that adding more virtual points (i.e., increasing the smoothing budget, $\alpha$) saves more query times. LIPP and SALI tend to perform quite similarly due to SALI using LIPP as the base index. For LIPP and SALI indexes, the easy to learn datasets (Facebook and Covid) stabilise after a certain number of virtual points are inserted. This is because the original datasets' CDFs are already quite linear. The same pattern was not observed for ALEX, this is because ALEX has an additional leaf-node search step not required by LIPP and SALI  (\algo\ forms larger nodes that could lead to longer leaf-node search times).

Fig.~\ref{fig:relative} further reports the average query improvements as a percentage against the original index structure over the promoted data.
It shows that applying \algo\ yields a query time improvement of up to 34\%, with stronger benefits observed over the two SOTA index structures LIPP and SALI. Smaller performance gain is observed over ALEX due to its leaf-node search process.  It is important to note that \algo\ still yields consistent query time improvements in this case. 
As $\alpha$ increases, the relative query time improvements for ALEX also increases, as the leaf-node search efficiency is improved due to increased accuracy of the refitted indexing functions. Since there is no such searching in LIPP and SALI's query process, their query performance is stagnant and the improvement represents the reduction in the index traversal time for query processing. 

 \begin{figure*}[t]
  \centering
  \captionsetup[subfigure]{justification=centering}  
  \begin{subfigure}[b]{0.31\textwidth}
    \includegraphics[width=\textwidth]{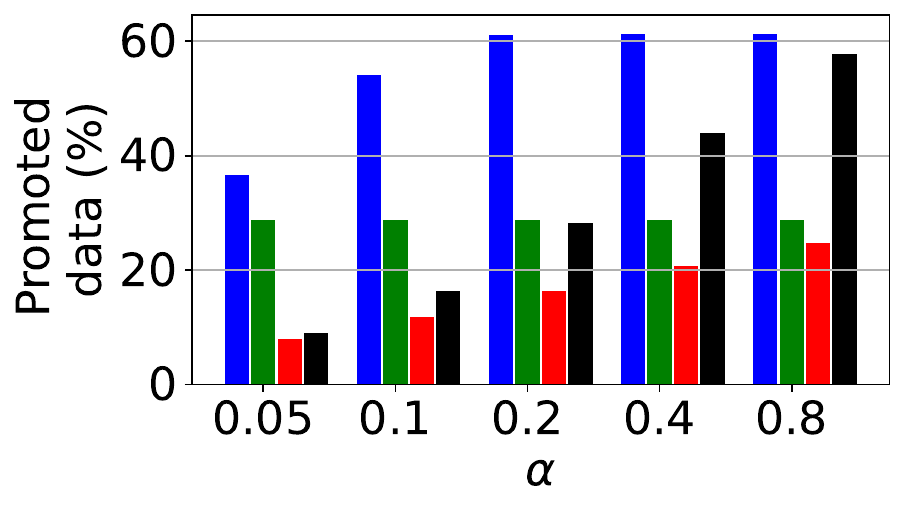}
    \caption{Percentage of promoted \\ data (LIPP)}
 \label{fig:promoted_data_lipp}
  \end{subfigure}
  \hfill
  \begin{subfigure}[b]{0.31\textwidth}
    \includegraphics[width=\textwidth]{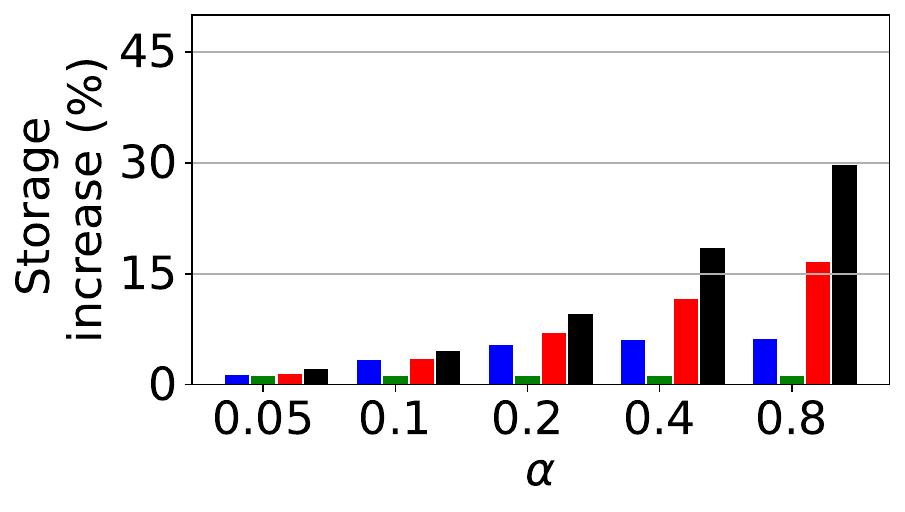}
    \caption{Percentage of storage\\ space increase (LIPP)}
 \label{fig:storage_change_lipp}
  \end{subfigure}
  \hfill
  \begin{subfigure}[b]{0.31\textwidth}
    \includegraphics[width=\textwidth]{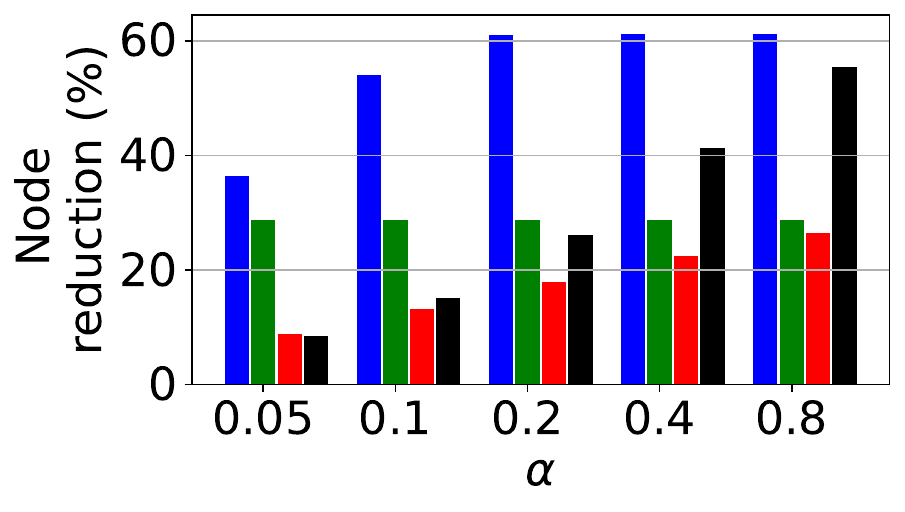}
    \caption{Percentage of index \\model reduction (LIPP)}
 \label{fig:node_change_lipp}
  \end{subfigure}\\
  
  \begin{subfigure}[b]{0.31\textwidth}
    \includegraphics[width=\textwidth]{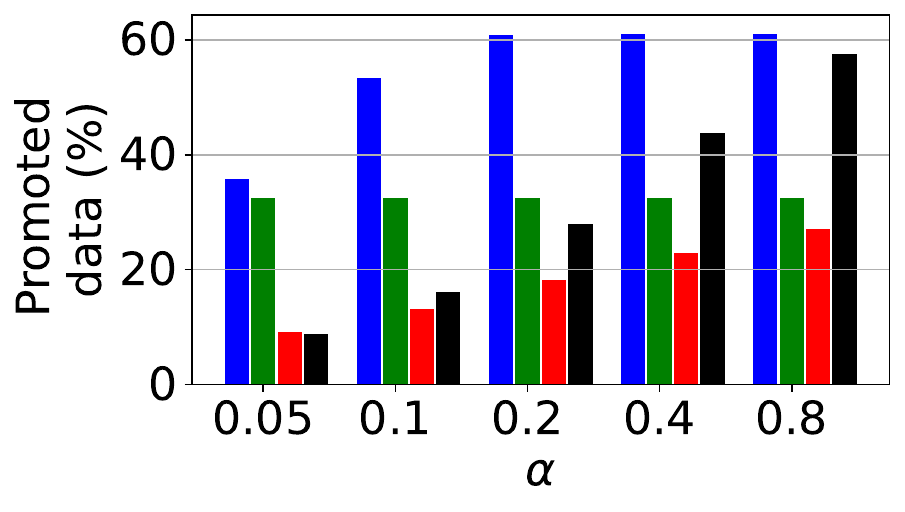}
    \caption{Percentage of promoted \\ data (SALI)}
 \label{fig:promoted_data_sali}
  \end{subfigure}
  \hfill
  \begin{subfigure}[b]{0.31\textwidth}
    \includegraphics[width=\textwidth]{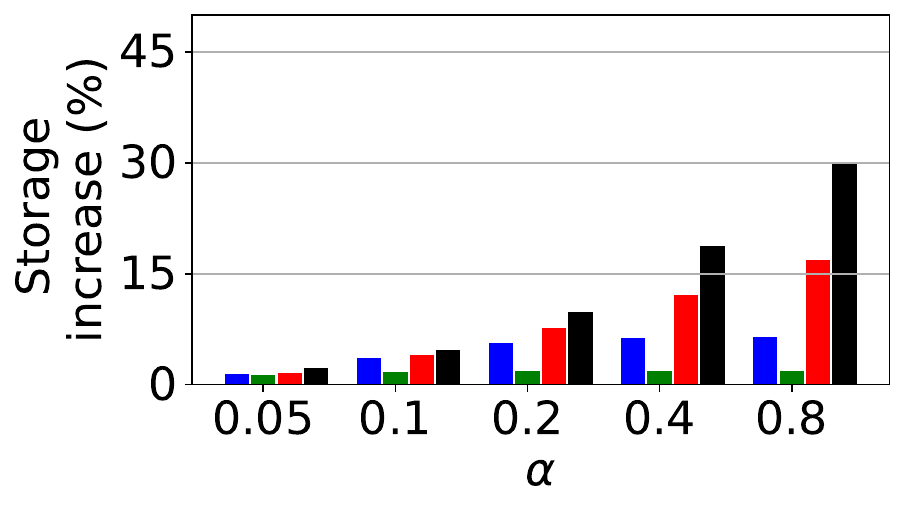}
    \caption{Percentage of storage\\ space increase (SALI)}
 \label{fig:storage_change_sali}
  \end{subfigure}
  \hfill
  \begin{subfigure}[b]{0.31\textwidth}
    \includegraphics[width=\textwidth]{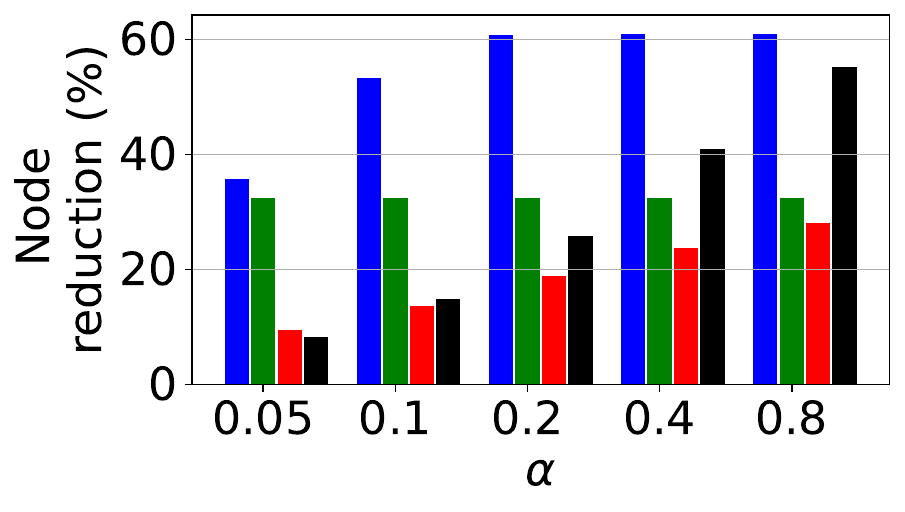}
    \caption{Percentage of index \\model reduction (SALI)}
 \label{fig:node_change_sali}
  \end{subfigure}\\

  \begin{subfigure}[b]{0.31\textwidth}
    \includegraphics[width=\textwidth]{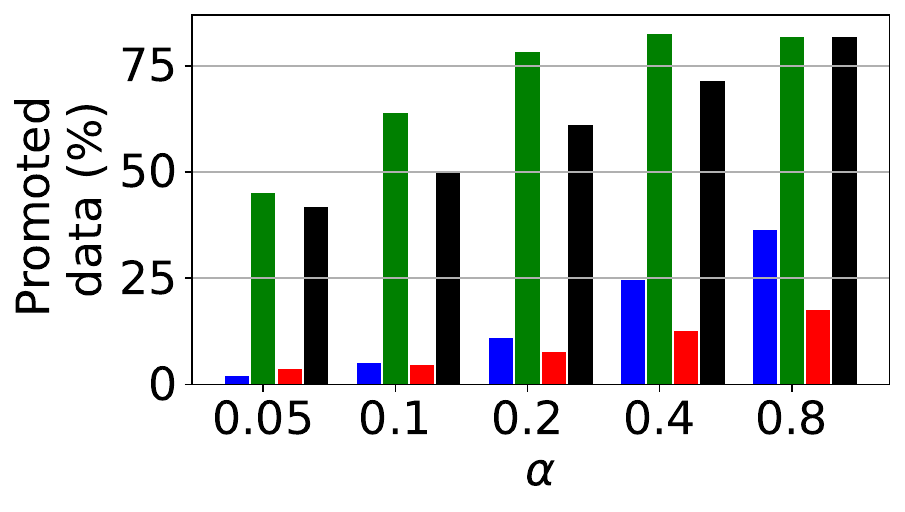}
    \caption{Percentage of promoted \\ data (ALEX)}
 \label{fig:promoted_data_alex}
  \end{subfigure}
  \hfill
  \begin{subfigure}[b]{0.31\textwidth}
    \includegraphics[width=\textwidth]{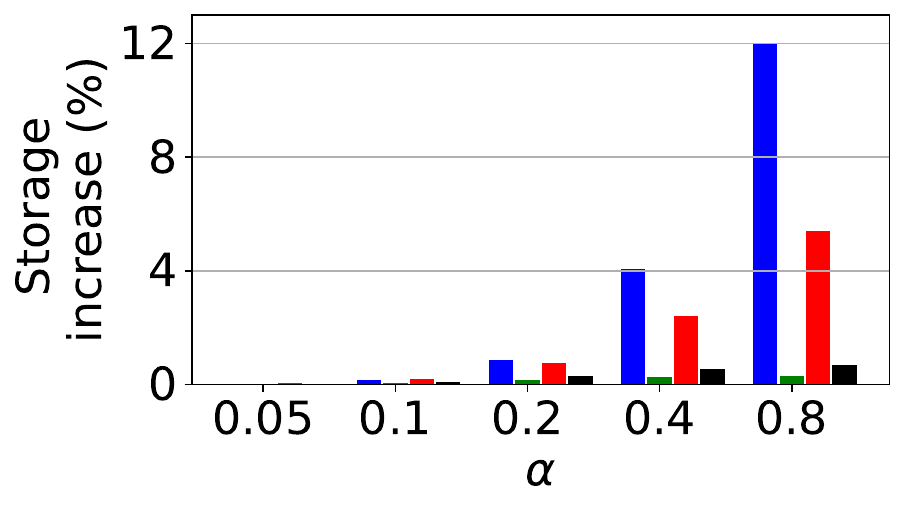}
    \caption{Percentage of storage\\ space increase (ALEX)}
 \label{fig:storage_change_alex}
  \end{subfigure}
  \hfill
  \begin{subfigure}[b]{0.31\textwidth}
    \includegraphics[width=\textwidth]{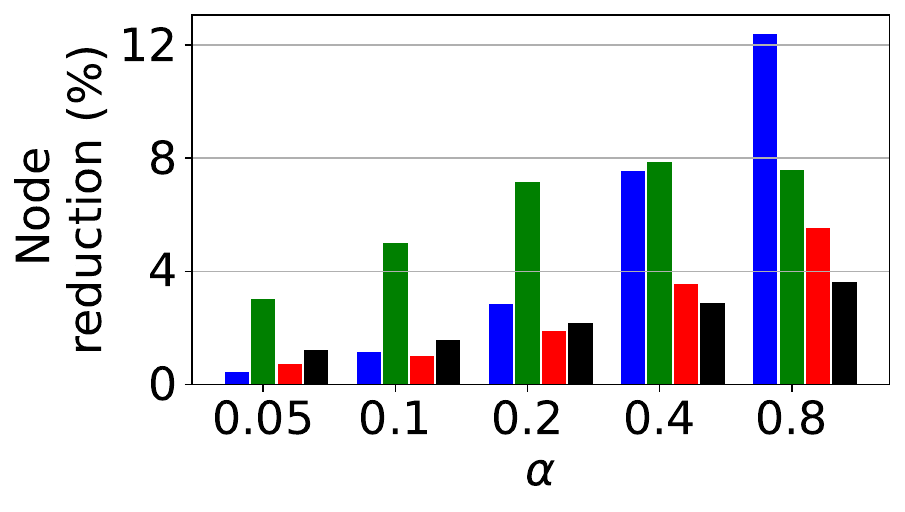}
    \caption{Percentage of index \\model reduction (ALEX)}
 \label{fig:node_change_alex}
  \end{subfigure}
  \caption{Space cost vs. smoothing threshold $\alpha$}
 \label{fig:space_alex}
\end{figure*}

\textbf{Size of the promoted data.} 
Figs.~\ref{fig:promoted_data_lipp},~\ref{fig:promoted_data_sali}, and~\ref{fig:promoted_data_alex} show the percentage of keys promoted.
For the Facebook dataset, \algo\ can promote around 60\% of all promotable data (i.e., keys at level 3 or below of the original index), while for the Covid dataset, \algo\ promotes around 30\% of the promotable data. For the harder to learn datasets, OSM and Genome, \algo\ also manages to promote up to 27\% and 57\% of the promotable data, respectively. The datasets with the most promoted data are again different for ALEX, due to its structural difference. 
Even though OSM and Genome report lower percentages of promoted data, the actual number of promoted keys is higher. This is because they have much more promotable data (i.e., more keys in lower levels).  Overall, as the smoothing threshold increases, more keys get promoted to upper levels. This is consistent with the theoretical analysis as adding more virtual points would allow more keys to be placed in nodes in upper levels. The difficult to learn key sets (OSM and Genome) demonstrate this property the most. This can also be matched with Fig.~\ref{fig:total_saved}, where the total time saved for OSM and Genome is higher due to higher number of data being promoted for those datasets, compared to the other datasets.

\textbf{Index size.} 
Due to the addition of virtual points, we expect the storage space consumption to increase. This is reflected in Figs.~\ref{fig:storage_change_lipp},~\ref{fig:storage_change_sali}, and~\ref{fig:storage_change_alex}. In most cases, less than 10\% of additional storage space is required by the \algo-enhanced indexes compared to the original structures, and in the worst case, less than 31\%. The space cost overhead is proportional to $\alpha$, which is also intuitive.

The storage space increase is balanced by the removal of unnecessary nodes, whose data is promoted to higher levels. 
Figs.~\ref{fig:node_change_lipp},~\ref{fig:node_change_sali},  and~\ref{fig:node_change_alex} report the node reductions achieved by \algo, which are given as the percentage of nodes that are removed relative to the number of nodes at levels 3 or lower of the original indexes. 
They follow similar patterns to the percentage of promoted data as expected. 

\textbf{Pre-processing time for \algo.}
The times taken to run \algo\ to optimise the learned index structures are summarised in Tables~\ref{tab:pre_lipp} and~\ref{tab:pre_alex} for LIPP and ALEX. As SALI is based on LIPP, where \algo\ reports a similar performance, we omit its results for brevity. 

\algo\ takes more time to run as $\alpha$ grows, which is consistent with our time complexity analysis. 
The algorithm running times vary across different datasets, again because the datasets have different difficulties in index learning. While the running times of \algo\ may seem quite large under certain settings, these are one-off pre-processing costs that can be amortised by the improvements in query time. One may further mitigate the impact of the extra construction times by using the original indexes for queries while constructing a parallel index structure with \algo. Once the \algo-optimised structure is ready, it is switched on for query processing.

\begin{table}[t]
\caption{\algo\ Pre-processing Time (s) for LIPP}
    \label{tab:pre_lipp}
    \centering
\begin{tabular}{lrrrrr}
\toprule
   $\alpha$        & 0.05  & 0.1  & 0.2 & 0.4 & 0.8 \\ \midrule
Facebook & 589       & 1,194 & 1,859 & 2,106 & 2,228  \\ \hline
Covid  & 304 & 337 & 343 & 337 & 336 \\ \hline  
OSM  & 1,217      & 2,329   & 4,495 & 7,983 & 13,019 \\ \hline
Genome  & 1,155       & 2,174   & 4,616 & 9,316 & 15,709 \\ 
\bottomrule
\end{tabular}
\end{table}

\begin{table}[t]
\caption{\algo\ Pre-processing Time (s) for ALEX}
    \label{tab:pre_alex}
    \centering
\begin{tabular}{lrrrrr}
\toprule
   $\alpha$        & 0.05  & 0.1  & 0.2 & 0.4 & 0.8 \\ \midrule
Facebook & 247      & 889 & 4,123 & 17,508 & 48,737  \\ \hline
Covid  & 609 & 1,423 & 2,795 & 4,463 & 4,955 \\ \hline  
OSM  & 988      & 2,297   & 9,526 & 33,097 & 81,620 \\ \hline
Genome  & 1,356       & 2,902  & 6,253 & 8,854 & 9,777\\ 
\bottomrule
\end{tabular}
\end{table}

\begin{figure*}[t]
  \centering
  
  \captionsetup[subfigure]{justification=centering}
 \begin{subfigure}[b]{0.31\textwidth}
    \includegraphics[width=\textwidth]{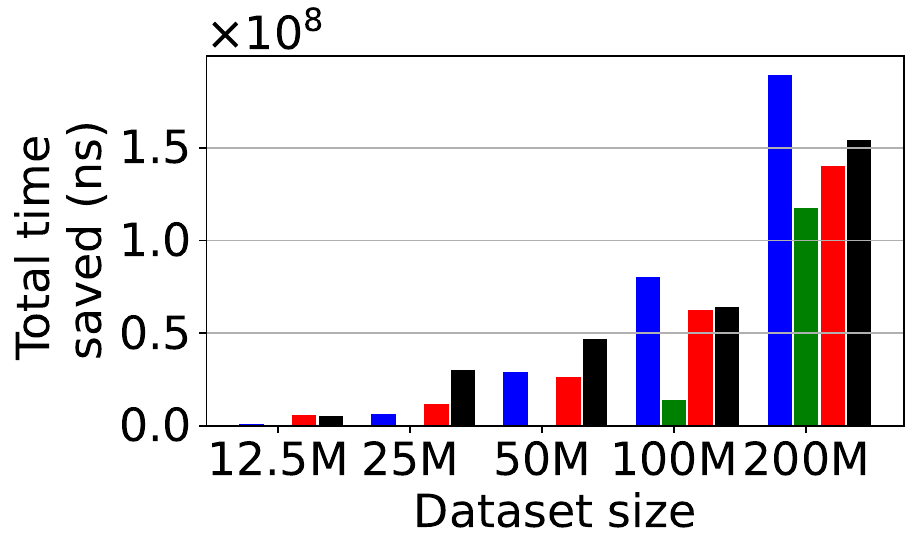}
    \caption{LIPP}
 \label{fig:scale_lipp}
  \end{subfigure}
  \hfill
  \begin{subfigure}[b]{0.29\textwidth}
    \includegraphics[width=\textwidth]{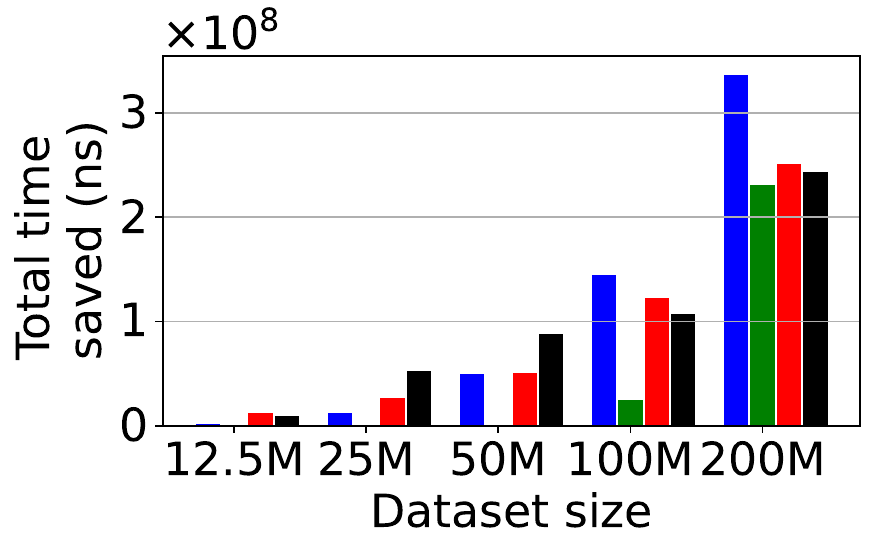}
    \caption{SALI}
 \label{fig:scale_sali}
  \end{subfigure}
  \hfill
  \begin{subfigure}[b]{0.29\textwidth}
    \includegraphics[width=\textwidth]{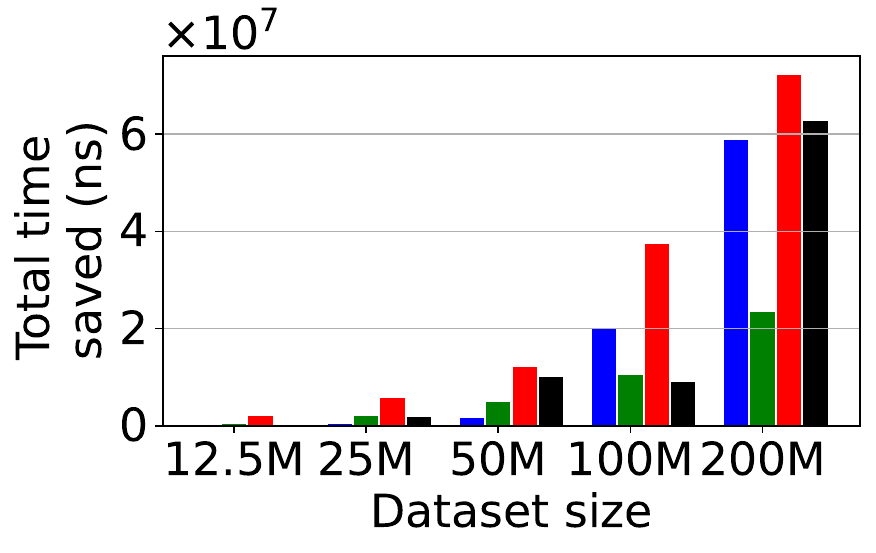}
    \caption{ALEX}
 \label{fig:scale_alex}
  \end{subfigure}
  \caption{Total time saved vs. dataset cardinality}
 \label{fig:cardinality}
\end{figure*}

\subsubsection{Impact of Dataset Cardinality.}
To demonstrate the scalability of \algo\ against the dataset cardinality, we repeat the experiments on datasets of 12.5 million to 200 million data points. Fig.~\ref{fig:cardinality} shows the total query times saved by applying \algo. For all datasets, the times saved grow with  the dataset cardinality, with faster growth being observed on the easier datasets (Facebook and Covid) grows faster. This is because there are not many keys in the lower levels for these datasets when the dataset cardinality is small.  These results confirm the scalability of \algo\ towards dataset cardinality.

\begin{figure*}[t]
  \centering
  
  \begin{subfigure}[b]{0.31\textwidth}
    \includegraphics[width=\textwidth]{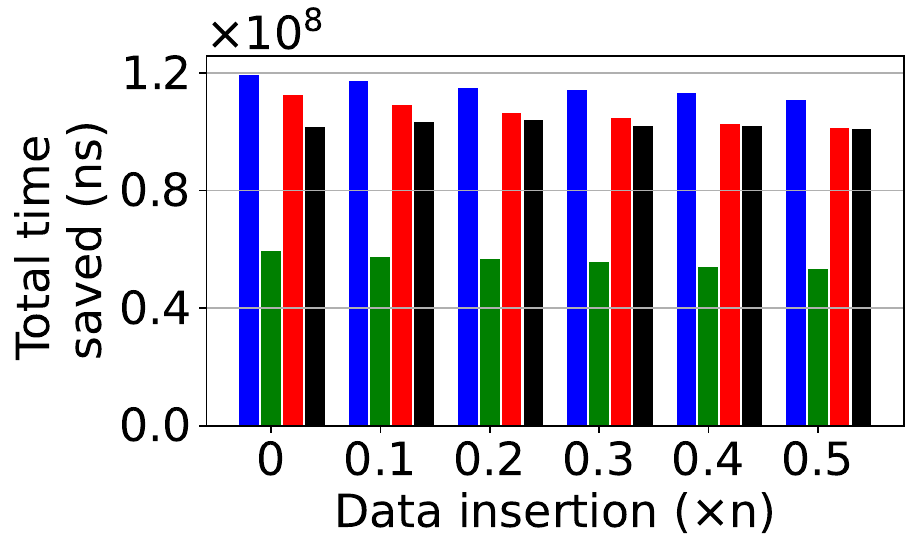}
    \caption{Total query time saved (LIPP)}
 \label{fig:query_improvement_lipp_insert}
  \end{subfigure}
  \hfill
  \begin{subfigure}[b]{0.31\textwidth}
    \includegraphics[width=\textwidth]{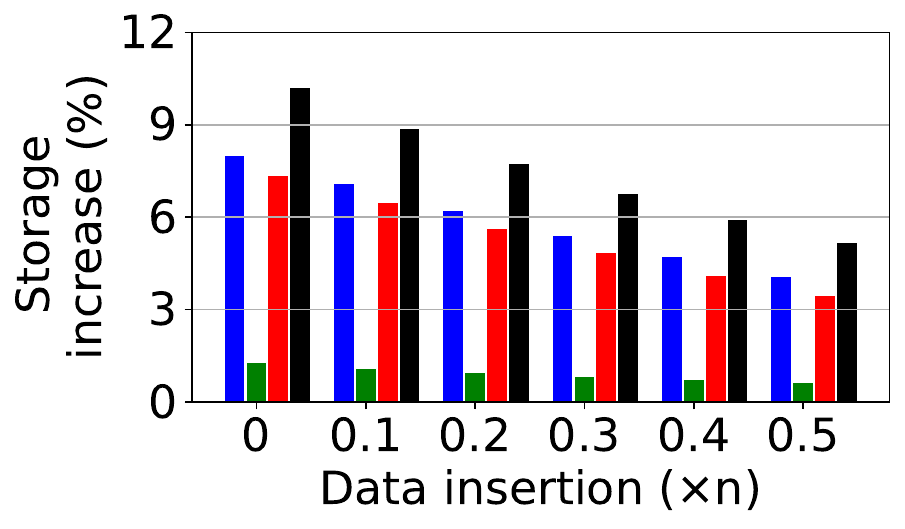}
    \caption{Storage increase (LIPP)}
 \label{fig:storage_change_insert_lipp}
  \end{subfigure}
  \hfill
  \begin{subfigure}[b]{0.31\textwidth}
    \includegraphics[width=\textwidth]{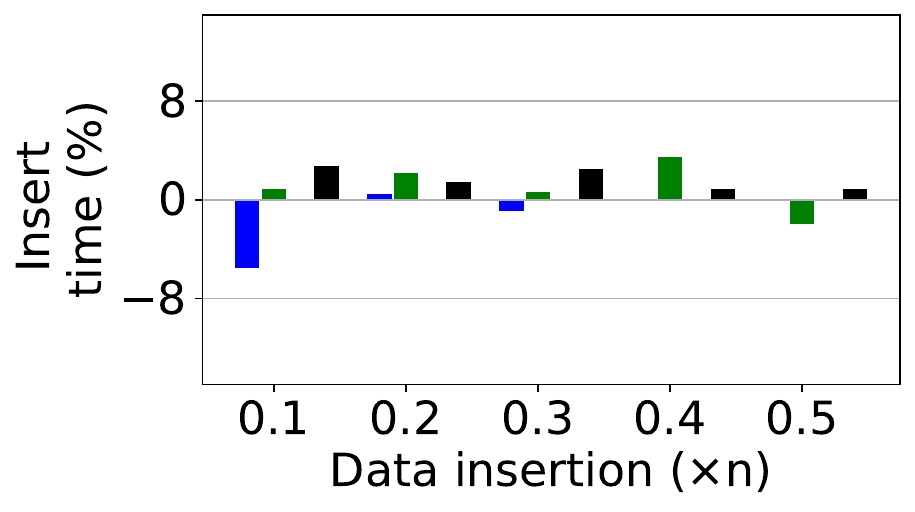}
    \caption{Insertion time increase (LIPP)}
 \label{fig:insert_lipp_insert}
  \end{subfigure}\\
  
  \begin{subfigure}[b]{0.31\textwidth}
    \includegraphics[width=\textwidth]{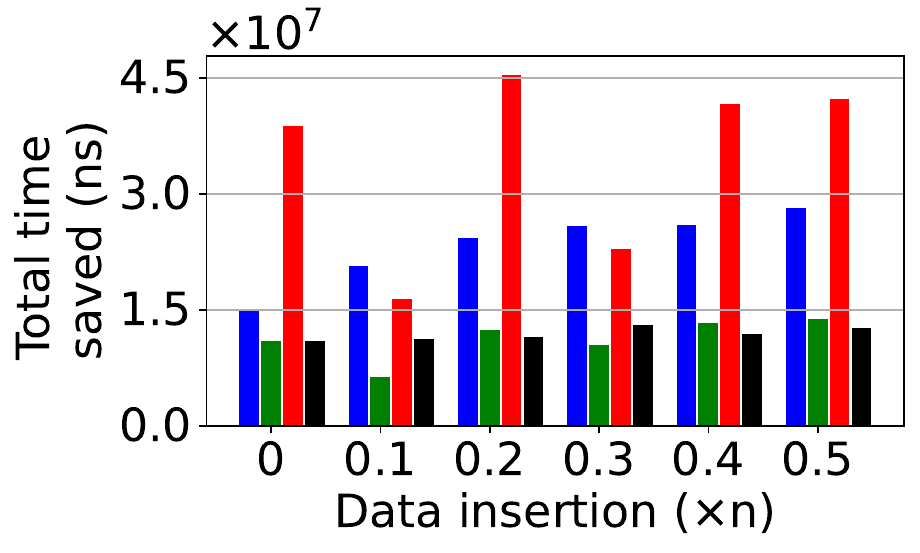}
    \caption{Total query time saved (ALEX)}
 \label{fig:query_improvement_alex_insert}
  \end{subfigure}
  \hfill
  \begin{subfigure}[b]{0.31\textwidth}
    \includegraphics[width=\textwidth]{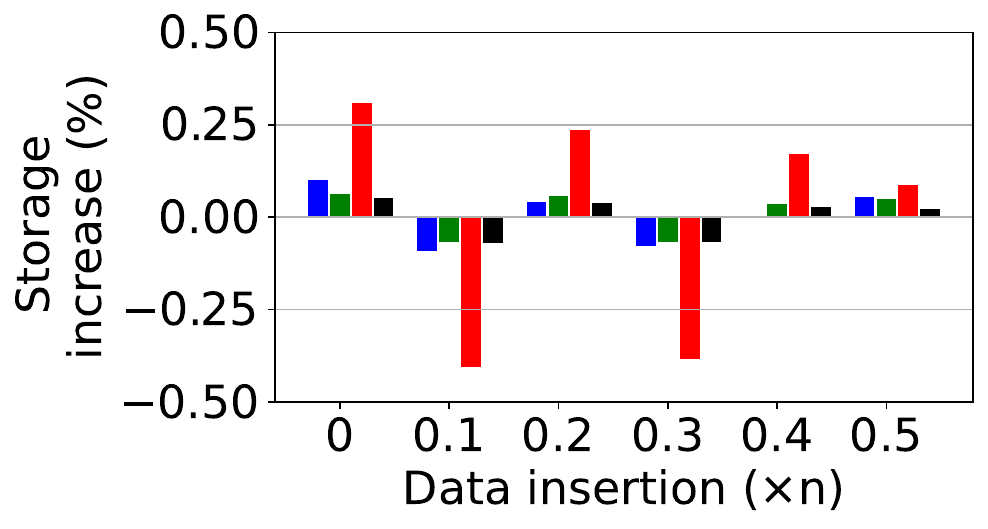}
    \caption{Storage increase (ALEX)}
 \label{fig:storage_change_insert_alex}
  \end{subfigure}
  \hfill
  \begin{subfigure}[b]{0.31\textwidth}
    \includegraphics[width=\textwidth]{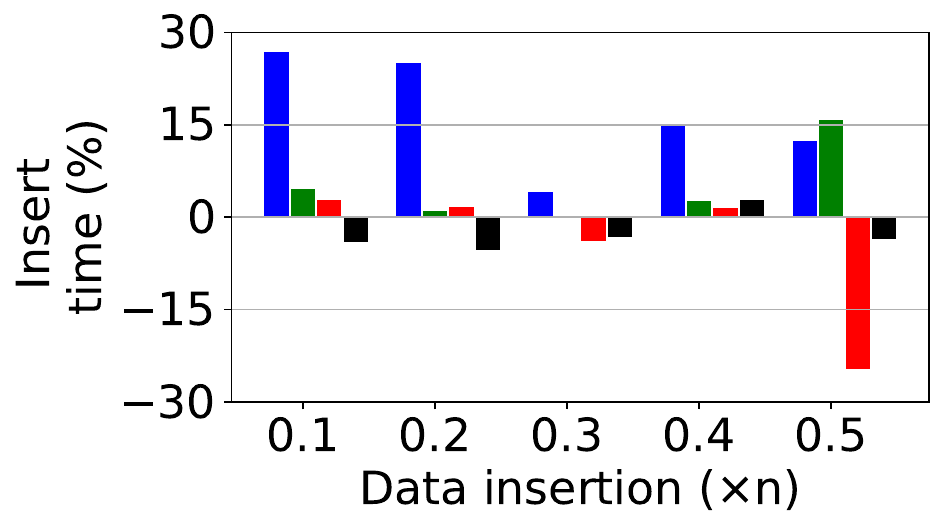}
    \caption{Insertion time increase (ALEX)}
 \label{fig:insert_alex_insert}
  \end{subfigure}
  \caption{Performance results  vs. data insertions}
 \label{fig:alex_read_write}
\end{figure*}

\subsection{Results on Read-write Workloads}

Due to the similar trends between LIPP and SALI indexes, SALI is omitted from the results below for brevity. The default smoothing threshold of 0.1 was used for these experiments.

\textbf{Query time (promoted data).}
Figs.~\ref{fig:query_improvement_lipp_insert} and~\ref{fig:query_improvement_alex_insert} show the total query times saved by \algo\ for LIPP and ALEX, respectively, compared to the original index structures, as more batches of data are inserted (recall that each batch consists of $0.1n$ data points). Here, the query times saved are decreasing slightly as more data points are inserted for LIPP, because the inserted data points have a higher chance of colliding with the promoted data points which are now in the upper levels, compared to when they are in lower levels as in the original index structure. For ALEX, the trend is quite similar except for on the OSM dataset, where there are two drops after one and three insertion batches (i.e., $0.1n$ and $0.3n$ data points are inserted). This is because the original index structure's query times happen to be slightly lower in these two cases.

\textbf{Index size.}
The index size overhead decreases after each  batch of insertions, as shown in Figs.~\ref{fig:storage_change_insert_lipp} and~\ref{fig:storage_change_insert_alex}. This is because the initial gaps left by the virtual points are gradually filled up by the inserted points, hence improving the overall space utilisation. Again, the index size overhead is at or below 10\%, emphasising the space efficiency of \algo. For ALEX index, the storage increase is almost negligible ($< 0.5\%$). In some cases, the storage size of the \algo-enhanced ALEX is even lower than the original index, as the original ALEX index may need to create more new nodes to host the insertions which outweighs the space overhead of \algo.

\textbf{Insertion time.}
Figs.~\ref{fig:insert_lipp_insert} and~\ref{fig:insert_alex_insert}  show the average insertion  times of the \algo-enhanced indexes relative to those of the original indexes. Using \algo\ helps improve the insertion times in some cases because the gaps left by the virtual points are reused for insertions. \algo\ could also lead to higher insertion times in other cases. 
This could be attributed to the fact that there are more keys at the upper levels of the \algo-enhanced indexes which may lead to more collisions with the insertions, which requires new index node creation. Overall, the insertion times of the \algo-enhanced indexes are on par to the original indexes.

\section{Conclusion}

We addressed the issue of index learning over data of complex distributions by a CDF smoothing technique to modify the key set, instead of developing yet another indexing function or structure. We proposed an algorithm named \algo\ to utilize this technique on existing hierarchical learned index structures, to improve the query time for the keys in lower levels of these index structures. The proposed algorithm is implemented on three recent learned indexes, which are evaluated on real-world datasets. The results show significant query performance improvements, i.e., up to 34\%, with a controllable and low storage space overhead. 

\bibliographystyle{ACM-Reference-Format}
\bibliography{references}

\end{document}